\documentclass{aa}  
\usepackage{graphicx}
\usepackage{txfonts}
\usepackage{longtable}
\begin{document}
   \title{Evidence for a Type~1/Type~2 dichotomy in the correlation between quasar optical
   polarization\\
          and host galaxy/extended emission position angles\thanks{Table 6 is only available in electronic form at the CDS via anonymous ftp to cdsarc.u-strasbg.fr.}}

      \subtitle{}

   \author{B. Borguet\inst{1}\fnmsep\thanks{PhD. grant student of the Belgian National Fund for Scientific Research (F.N.R.S.)}
          \and D. Hutsem\'ekers\inst{1}\fnmsep\thanks{Senior research associate F.N.R.S.}
          \and G. Letawe\inst{1}
          \and Y. Letawe\inst{1}
          \and P. Magain\inst{1}
          }

   \offprints{B. Borguet}

   \institute{Institut d'Astrophysique et de G\'eophysique, University of Li\`ege,
              All\'ee du 6 Ao\^ut 17, B-4000 Li\`ege\\
              \email{b.borguet@ulg.ac.be}
             }

   \date{Received ; accepted }

%
  \abstract
%
   {}
%
%
     {For Seyfert galaxies, the AGN unification model provides a simple and well established
      explanation of the Type~1/Type~2 dichotomy through
      orientation based effects.
     The generalization of this unification model to the higher
     luminosity AGNs that are the quasars remains a key question. The recent
     detection of Type~2 Radio-Quiet quasars seems to support such an extension.
     We propose to further test this scenario.}
%
   {On the basis of a compilation of quasar host galaxy position
   angles consisting of previously published data and of new measurements
   performed using HST Archive images, we investigate the possible existence of a correlation between
   the linear polarization position angle and the host galaxy/extended emission position angle of quasars.}
   {We find that the orientation of the rest-frame UV/blue extended emission is correlated
   to the direction of the quasar polarization. For Type~1 quasars, the polarization is aligned with the
   extended UV/blue emission while these two quantities are perpendicular in Type~2 objects. This result is
   independent of the quasar radio-loudness. We interpret this (anti-)alignment effect in terms of scattering
   in a two-component polar+equatorial model which applies to both Type~1 and Type~2 objects. Moreover the orientation
   of the polarization --and then of the UV/blue scattered light-- does not appear correlated to the major axis
   of the stellar component of the host galaxy measured from near-IR images.}

   {}

   \titlerunning{Correlation between optical polarization and host morphology in quasars.}
   \authorrunning{B. Borguet et al.}

   \maketitle
%

\section{Introduction}

   The study of quasars shows that we can classify them among various
   categories. Radio-Loud quasars (RLQ) are distinguished from the Radio-Quiet quasars (RQQ) according to their radio
   power (Kellerman et al. \cite{ke89}), and the
   Type~1/Type~2 objects from the presence or absence of broad emission lines in
   their spectrum (Lawrence \cite{law87}).

   One can then wonder whether there exists a common link between
   all these objects, i.e. are the physical processes at the
   origin of all quasars the same? An interesting way to tackle this question
   consists in the use of the linear optical polarization.
   Polarimetry, in combination with spectroscopy, has
   led to major advances in the development of a unified scheme
   for AGN (see Antonucci \cite{anto93} for a review). The discovery of hidden broad emission lines in the
   polarized spectrum of the Type~2 Seyfert galaxy NGC1068
   led to consider these objects as intrinsically identical to Type
   1 Seyfert, the edge-on orientation of a dusty torus blocking
   the direct view of the central engine and the broad
   emission line region (Antonucci \& Miller \cite{anto85}).

   The question we investigate in this paper relates to
   the possible existence of a correlation between the optical polarization position angle $\theta_{Pola}$\footnote{The polarization position angle $\theta_{Pola}$ is defined as the position
   angle of the maximal elongation of the electric vector in the plane of polarization, measured in degrees East of North.}
   and the orientation of the host galaxy/extended emission $PA_{host}$\footnote{The orientation of the host galaxy $PA_{host}$ is characterized by the position
   angle of its major axis projected onto the plane of the sky, measured in degrees East of North.} in the case of
   RLQs and RQQs.
   The relation between quasars and their host galaxies may play a fundamental role in our understanding of the AGN phenomenon and in determining the importance
   of the feedback of AGN on their hosts.

   Such a correlation has already been studied in the case of the less
   powerful AGN that are the Seyfert galaxies. Thompson \& Martin
   (\cite{toma88}) found a tendency for Seyfert 1 to have their
   polarization angle aligned with the major axis of their host galaxy,
   an observation that they interpreted as due to dichroic extinction by aligned dust grains in the
   Seyfert host galaxy. In the case of the more powerful AGN that are
   quasars, Berriman et al. (\cite{beri90}) investigated
   this question. They determined by hand
   the $PA_{host}$ of 24 PG quasars from ground based images
   and computed the acute angle $\Delta \theta$ between $\theta_{Pola}$ and
   $PA_{host}$. They observed that while more objects seem to appear at small
   values ($\Delta \theta \leq 45\degr$), this effect was only
   marginally statistically significant.

   Investigating this problem with ground based data for Type~1 quasars is
   hampered by the inability of separating the faint
   host galaxy/extended emission\footnote{In the following, we use indifferently the term host
   galaxy or extended emission to refer to all the extended emission around the central source
   including stars, ionized gas or scattered light.} from the blinding light of the quasar
   nucleus. Because of its high angular resolution and stable Point Spread Function (PSF),
   the Hubble Space Telescope (HST) permits to properly remove the
   contribution of the powerful quasar nucleus thus allowing the investigation of the host galaxy parameters.
   A large number of quasar host observing programs
   were carried out with the HST leading to a statistically useful sample of quasar host
   galaxy parameters (e.g. Bahcall et al. \cite{ba97}; Dunlop et
   al. \cite{du03} and many more see Sects.~\ref{publidata} and \ref{newosdat}).
   Our aim is to investigate the $\theta_{Pola}/PA_{host}$
   relation on the basis of high resolution HST quasar images. We
   will use either position angles given
   in the literature or $PA_{host}$ we measured ourselves from HST Archive observations
   (hereafter called ``\emph{new $PA_{host}$ data}").

   The layout of this paper is as follows. In Sect.~\ref{publidata} we
   introduce the samples of quasars used in this study
   which possess a $PA_{host}$ given in the literature.
   In Sect.~\ref{tres}, we outline the samples with good imaging data but for which no $PA_{host}$
   were published, and we summarize our data
   analysis process and the approach followed to model the HST Archive images
   and to derive the host galaxy parameters. In Sect.\ref{rapola} we briefly
   describe the polarization and radio data. Then in Sect.~\ref{statos} we
   present the statistical analysis of the sample and the
   results obtained. In Sect.~\ref{discu} we discuss the results and
   compare them to former studies. Finally, our conclusions are
   summarized in Sect.~\ref{conclu}.


\section{Description of the sample : published data}

 \label{publidata}

   In this section we give details about the observing campaigns
   used in the present study excepting the \emph{new $PA_{host}$ data} which will be
   discussed in the next section.
   Note that our study is essentially based on
   intrinsically high luminosity Radio-Loud and Radio-Quiet AGN (i.e. quasars), but
   does not consider BLLac objects since these objects
   show highly variable luminosity and polarization
   (e.g. Urry \& Padovani \cite{urry95}). We also select the Type~2 counterpart of Radio-Loud quasars
   (the Narrow Line Radio Galaxies, NLRGs, Barthel \cite{bar89}) from the whole RG
   sample on the basis of classifications given in the
   literature (Grandi et al. \cite{gr06}; Haas et al. \cite{haa05}; Cohen et al. \cite{coh99}; Rudy et al.
    \cite{rudy83}; Tadhunter et al. \cite{tad98}; Tadhunter et al. \cite{tad02}; van
    Bemmel \& Bertoldi \cite{vb01}; Siebenmorgen et al. \cite{si04} and Jackson \& Rawlings \cite{ja97}).

   In the following we present the
   samples which were analyzed in the literature and the
   methods used by authors to recover the host galaxy
   parameters from the quasar images.
   For more suitability in the
   following text, we take the convention to speak of the
   data of \emph{Ba97} to refer to the images/data published in the
   paper of Bahcall et al. (\cite{ba97}), and so on for the other
   papers.


Note that in our final sample, we only consider objects from these
samples if they actually possess both a reliable host position
angle and an accurate enough polarization angle. So only
part of the objects presented in the papers belongs to our final
sample. A detailed description of the observations and $PA_{host}$
measurements is available in Appendix~\ref{reli}.

\paragraph{The visible domain :}

  this sample is constituted of data published in the
  following papers, chronologically ordered : Cimatti et al. \cite{cima93}, Disney et al. \cite{di95},
  de Koff et al. \cite{dk96}, Lehnert et al. \cite{le99}, Martel et al. \cite{ma99},
  Dunlop et al. \cite{du03}, Floyd et al. \cite{fl04}
  and McLure et al. \cite{mc04}, respectively abbreviated \emph{Ci93}, \emph{Di95},
  \emph{Dk96}, \emph{Le99}, \emph{Ma99}, \emph{Du03}, \emph{Fl04}
  and \emph{Mc04}.

  The radio galaxy and quasar images studied in these papers were mostly (except the RGs from \emph{Ci93}) imaged
  through an R-band filter with the HST WFPC2 camera. The sample spans a
  redshift range from $z =$ 0.0 to $z =$ 2.0 and is essentially made out of 3CR and PG
  objects. Considering these samples, we obtain a final number of
  79 objects possessing a published $PA_{host}$ in the
  visible domain together with a polarization position angle.

\paragraph{The near-IR domain :}

  the near-IR compilation contains several samples coming from
  surveys using the NICMOS HST camera or ground based images.
  All the targets of the resulting sample were observed in the near-IR using K and H-band filters. The
  host galaxy data were retrieved from the papers of Taylor et al. \cite{ta96}, Percival et
  al. \cite{pe01}, Guyon et al. \cite{gu06} and Veilleux et al.
  \cite{ve06} abbreviated respectively \emph{Ta96}, \emph{Pe01},
  \emph{Gu06} and \emph{Ve06}. The compiled resulting sample, considering
  only the objects for which we have at least a $PA_{host}$ and a polarization position angle contains 21
  objects with  redshifts $z \le 0.4$.

\paragraph{Determination of the $PA_{host}$ :}
  \label{hostdet}

  the $PA_{host}$ published in the aforementioned papers were measured
  as follows.
  The first step to determine the position angle of a quasar host galaxy is the
  proper subtraction of the bright central source which dominates the optical emission of Type~1 objects.
  To this aim, a careful PSF construction is realized in order to precisely characterize
  the profile of a point like source imaged by a given instrument. This PSF is then scaled to
  the quasar intensity and subtracted from the quasar image allowing
  in most cases the detection of the underlying host.
  This first step is realized for all quasars except for Type~2
  objects in which the heavy obscuration of the central source
  provides a non-contaminated view of the host.

  Several techniques are then used to derive the host
  parameters ($PA_{host}$ and $b/a$) either ellipse fitting
   on the host image followed by the measurement of the model position
  angle at a chosen surface brightness, or two-dimensional galaxy profile fitting over the nucleus-subtracted
  image. These techniques give rather similar results since the
  crucial step remains the proper separation between the point-like component
  and the underlying host.



\section{Measurements of new $PA_{host}$}
\label{tres}

\subsection{The samples}
\label{newosdat}

  The samples are constituted of available
  quasar/RG HST observations retrieved from the Archive, but for which the images and more particularly
  the host galaxy parameters were not thoroughly studied or published in
  the literature. Detailed observational properties of these samples are available in
  Appendix~\ref{newly}, while major properties are summarized in Table~\ref{tableparam}.

\paragraph{The visible domain :}

  this sample gathers quasars imaged in the V to R-band
  using the HST WFPC2 or ACS camera. The total sample consists
  of 53 objects coming from the following papers:
  Bahcall et al. \cite{ba97}, Boyce et al. \cite{bo98}, Marble et
  al. \cite{ma03} and Zakamska et al. \cite{za06} hereafter
  abbreviated \emph{Ba97}, \emph{Bo98}, \emph{Ma03} and \emph{Za06} respectively.
  This sample essentially contains $z \le 0.5$ 3CR, 2-Micron All Sky Survey (2MASS), Sloan Digital Sky Survey
  (SDSS) and PG quasars.

\paragraph{The near-IR domain :}

  this sample is a compilation of 17 RQQs and RLQs from the McLeod et al.
  \cite{mc01} and Kukula et al. \cite{ku01} papers (hereafter abbreviated \emph{Ml01} and \emph{Ku01}). The objects,
  spanning a redshift range $0.1 < z  < 2.0$ were observed through the H or J-band
  by the NICMOS camera on board of HST.

\subsection{The method}
\label{newmeas}

   Here we describe the method we used to derive host galaxy position angle and axial ratio
   from the images.
   First we outline the
   images reduction steps and then we give an overview of the
   technique used to disentangle the quasar nuclear and host galaxy contributions. We then
   present the construction of the Point Spread Function (PSF) necessary for an accurate separation.
   The results of the modelling are finally given and compared to
   the results of other studies in the following subsection.


\begin{table}
\begin{minipage}[t]{\columnwidth}
\caption{Characteristics of the surveys used to derive new host
galaxy position angles.}
\label{tableparam}      
\centering
\renewcommand{\footnoterule}{}
\begin{tabular}{cccc}        
\hline \hline                 

Reference\footnote{Reference to
the paper where the images were published.} & Prop ID\footnote{Corresponding HST Proposal ID.} & Instrument\footnote{Name of the camera.} & Filter\footnote{Name of the filter used.} \\    

\hline                        

   \emph{Ba97} & 5343,5099 & WFPC2-WF3 & F606M \\      
   \emph{Bo98} & 6303,5143 & WFPC2-PC1    & F702M \\
   \emph{Ml01} & 7421 & NICMOS-NIC2     & F160M \\
   \emph{Ku01} & 7447 & NICMOS-NIC1     & F110M\&F165M \\
   \emph{Ma03} & 9057 & WFPC2-PC1     & F814M \\
   \emph{Za06} & 9905 & ACS-WFC    & F550M \\
\hline                                   

\end{tabular}
\end{minipage}
\end{table}
%

   \subsubsection{Image reduction}

   The quasar images were retrieved from the MAST HST Archive of the Space Telescope
   Science Institute (STScI, http://archive.stsci.edu/astro). All
   the images are already pre-processed by standard STSDAS pipeline
   softwares adapted to each HST instrument. This pre-processing
   essentially consists in dark frame and bias substraction
   along with flat-fielding, with the best reference files available.
   In the case of the ACS images, where the geometric distortion in the WFC camera is significant, an
   additional step was added to correct these distortions thanks
   to the $MultiDrizzle$ routine (Koekemoer et al. \cite{ko02}).

   In addition to this pre-processing, we carried out two
   additional processing steps. The first consists in
   the removal of cosmic rays using the $crrej$
   task of the Image Reduction and Analysis Facility (IRAF) software which
   rejects high pixels from sets of exposures of the same field in an iterative way. These bad pixels
   are not used in the creation of the output image providing a
   final cosmic ray free image. In the case of the NICMOS data,
   this step is not essential since the MULTIACCUM mode used in the image acquisition phase already
   realizes this processing when acquiring the data (Schultz et al. \cite{shu05}).

   The second step consists in the
   background subtraction. As the backgrounds on the WFPC2 and ACS WFC camera are
   extremely flat around the targets of interest
   we simply estimated the mean of the background level in each image from
   several subset of pixels excluding obvious sources, and then
   subtracted this mean value to each pixel of the image. In the case of
   the NICMOS data, the known pedestal effect (variable quadrant
   bias of the detectors) was not taken into account since the
   objects images were always almost centered in one of the quadrants.

   \subsubsection{Image processing}
     \label{decoprocess}

   As our primarily aim is to determine the position angle of the
   host galaxy surrounding nuclear dominated objects, we have to distinguish
   the faint fuzz of the galaxy from the quasar. The use of high resolution HST images partly
   solves this problem, but rises another one : the complex path
   of the light rays in the instrument create a complex PSF with quite extended wings that
   can cover, scaled to the intensity of the quasar, the underlying galaxy.
   The use of a deconvolution technique makes it possible to separate quite efficiently the light
   of the point like quasar central source (roughly a scaled PSF) from the light of the
   host. We used a version of the MCS algorithm (Magain et al.
   \cite{ma98}) that is well suited to deconvolve the HST images while
   adjusting an analytical galaxy model.

   Our image processing proceeds in essentially two steps. In the
   first stage, the parameters (intensity and position) of the point source are
   adjusted to subtract the contribution of the quasar.
   This contribution is estimated so that a minimal amount of PSF structure remains in the image.
   This step has to be realized with particular care as the inner
   pixels (out to a radius of 5 pixels for the extreme cases) of the quasar image
   may be saturated, so that the PSF subtraction process may become quite subjective.

   The second stage corresponds to the fitting of an analytical
   galaxy profile (properly convolved with the HST PSF) to the PSF subtracted images. The analytic profile we used consists in a typical S\'ersic
   (S\'ersic \cite{se68})
   profile:

                       \begin{equation}  \label{galaxmod}
                     {I(x,y) = I_{0} e^{-(A x^{2} + B y^{2} + C x y)^{\alpha}}}
                       \end{equation}
    where $I_{0}$ stands for the central surface brightness, and
    $\alpha$ describes its profile (the profile is called elliptical or \emph{de Vaucouleurs} if
    $\alpha = 1/8$ and spiral or disk-like if $\alpha = 1/2$).
    A coordinate change allows us to recover the axial
    ratio ($b/a$) and the position angle ($PA_{host}$) of the model fitted to the data.
    Since we are essentially interested in the orientation of the major light concentration, we only
    use in the following a \emph{de Vaucouleurs} profile, noting after some tests that $PA_{host}$
    does not strongly depend on $\alpha$.

    \subsubsection{PSF creation}

    The quality of the image processing critically depends on
    the accuracy of the PSF used. A careful construction of
    the PSF is of prime importance to properly separate
    the contribution of the bright nucleus from that of the host galaxy.
    In the MCS method adapted to HST images, the construction of the PSF proceeds in two steps.
    The first step consists in the fitting of a sub-sampled
    numerical estimate of the PSF computed with the TinyTim package (Krist \& Hook
    \cite{kr04}) to an original resolution one (also estimated with TinyTim).
    This step provides the \emph{Tinydec} i.e. a PSF that
    has a better resolution than the observed one.
    The second step consists in a second fit,
    where the \emph{Tinydec} is fitted to an
    observed PSF (typically a calibration star lying in a weakly crowded
    field) in order to improve the agreement between the numerical estimate and the observed point source.
    This second fit is important since the TinyTim package is
    unable to give an accurate estimation of the external fainter
    halo ($r > 2 \arcsec$) as well as of the diffraction spikes (McLure et al.
    \cite{ml00}; Pian et al. \cite{pi02}).

    Usually one or more stars were imaged during
    each of the considered campaigns with growing exposures times,
    allowing the construction of a deep unsaturated PSF star.
    The aforementioned procedure could not be used in the case of
    the \emph{Ma03} and \emph{Za06} campaigns since no
    time was devoted to the observation of a specific PSF star, then
    forcing us to use the \emph{Tinydec} as the PSF.
    However the lack of observed standard PSF star is less
    constraining in these two cases since the quasars imaged are generally heavily
    obscured objects such that the central source contribution and the intensity of the associated scattered halo are
    weak (or null for the Type~2 objects of \emph{Za06}).

    \subsubsection{The error frame}

    In order to determine whether a given quasar model provides an
    accurate description of an observed image, a $\chi^{2}$
    minimization is used by the MCS program :

    \begin{equation} \label{chistat}
      {\chi^{2}= \sum _{all pixels} [\frac{Image-(PointSource+Host)\otimes PSF}{\sigma}]^{2},}
   \end{equation}
    where $Image$ represents the reduced quasar image, $PointSource+Host$ the quasar model, and
    $\sigma$ the standard deviation in the corresponding pixel (called the error frame). A
    careful construction of the error frame is mandatory for each image in order to
    avoid biased results.

    The error term accounts for different elements : first, for a
    Poissonian component due to photon shot noise and dark current as well as the noise
    induced by the CCD readout. The second component accounts for
    the detector pixels that are known to work badly, these pixels
    being indexed in flag frames available via the STScI website.
    We used this frame to construct masks such that
    zero weight (i.e. $1/\sigma = 0$) is allocated to these indexed pixels while this value is set
    to one everywhere else over the frame. The last component
    accounts for saturated regions in the observed image due to the long exposure time required for the imaging of the
    faint host galaxies. We accounted for this final component
    by assigning a null weight to the implied pixels. These masks
    are then multiplied by the original $1/\sigma$ constructed in
    the first step.

    \subsection{The resulting $PA_{host}$}

    The results of our image analysis are summarized in Table~\ref{imres}.
    Columns are as follows : (1) object name,
    (2) redshift of the object, (3) radio-loudness of the object (RL or RQ), (4) source of
    the image, (5) spectral domain of the image (visible (Vis) or near-IR (NIR)), (6) type of the host
    galaxy (E : elliptical, S : spiral and U : undefined), (7) and (8) axial ratio
    and position angle of the host ($PA_{host}$ in degrees, East of
    North) and (9) a quality criterion (1, 2 or 3, with 1 being the highest quality). The parameters in Col.~(6) and
    Col.~(9) are defined in the following subsections.

\begin{figure*}
      \centering
   \includegraphics[width=4.4cm]{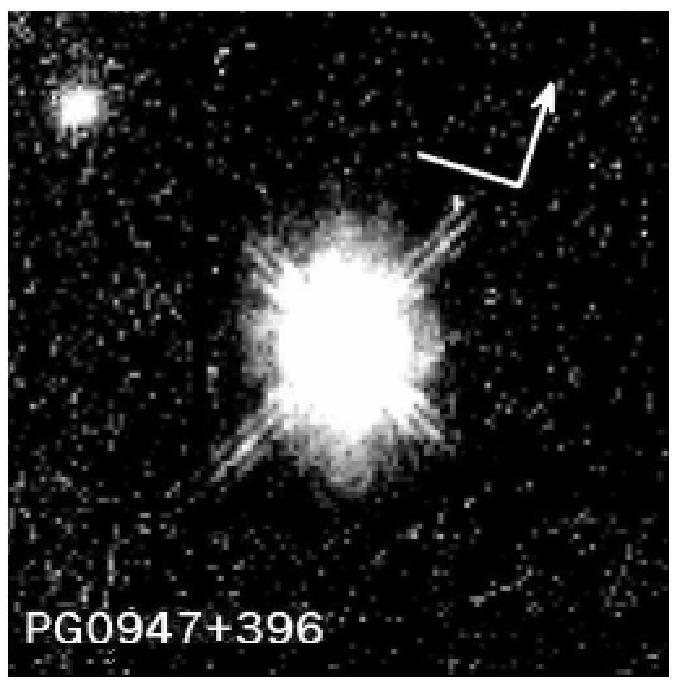}
   \includegraphics[width=4.4cm]{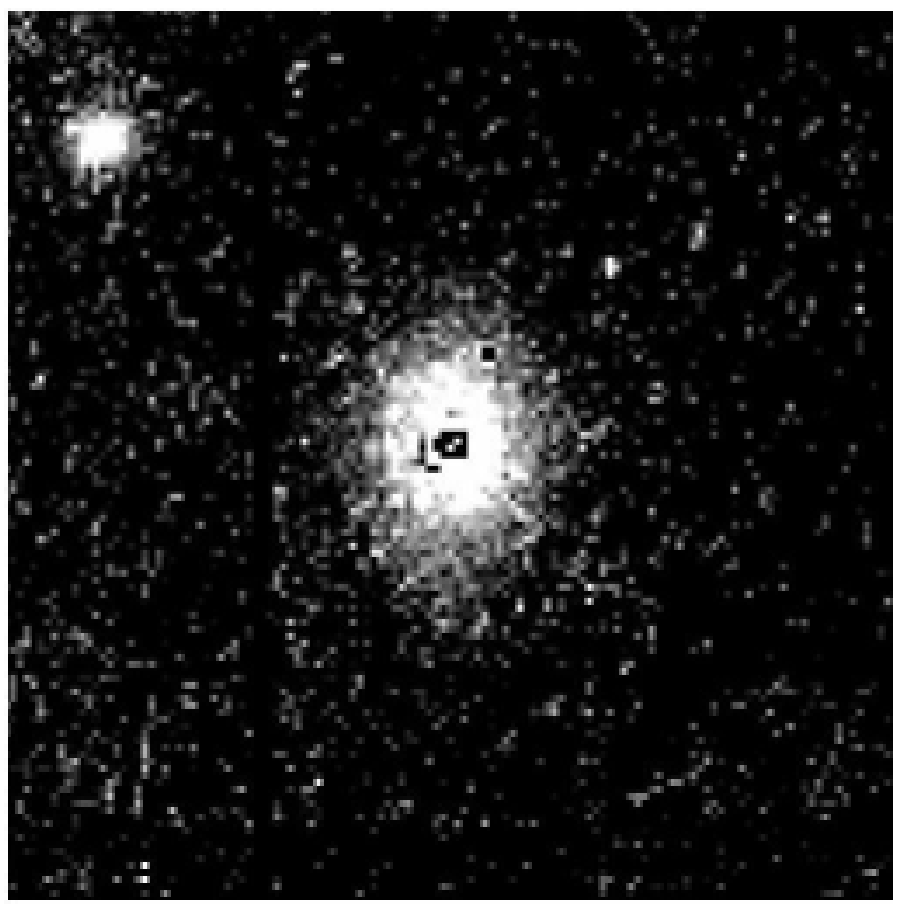}
   \includegraphics[width=4.4cm]{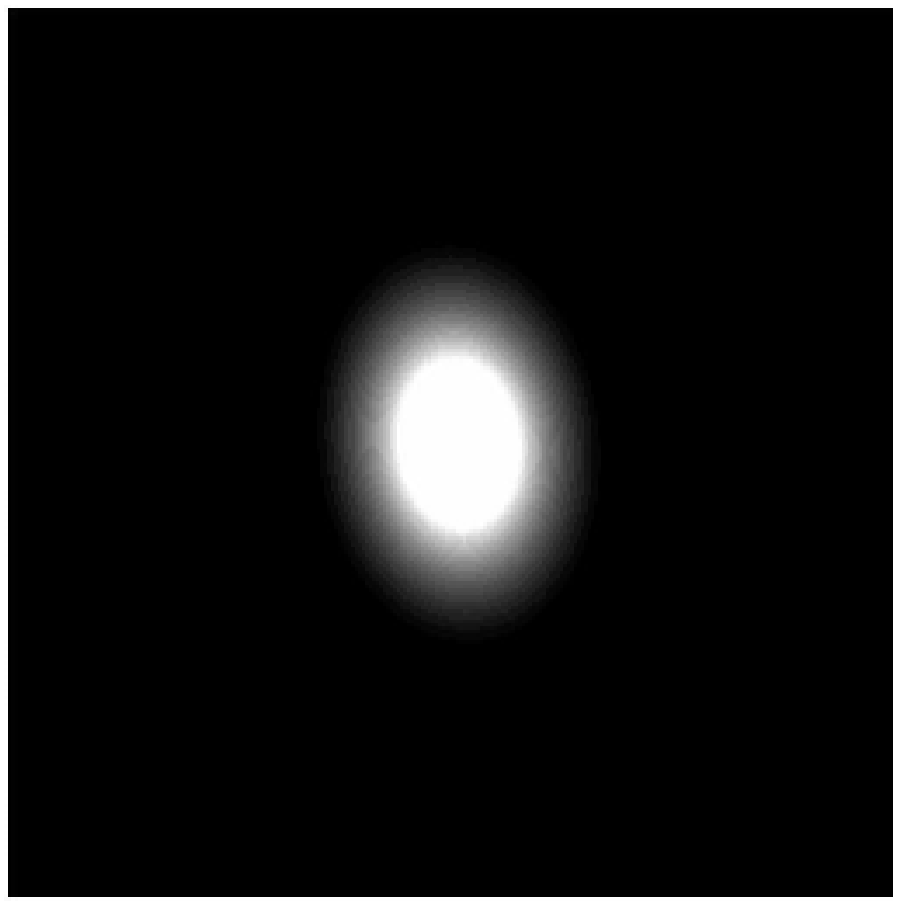}
   \includegraphics[width=4.4cm]{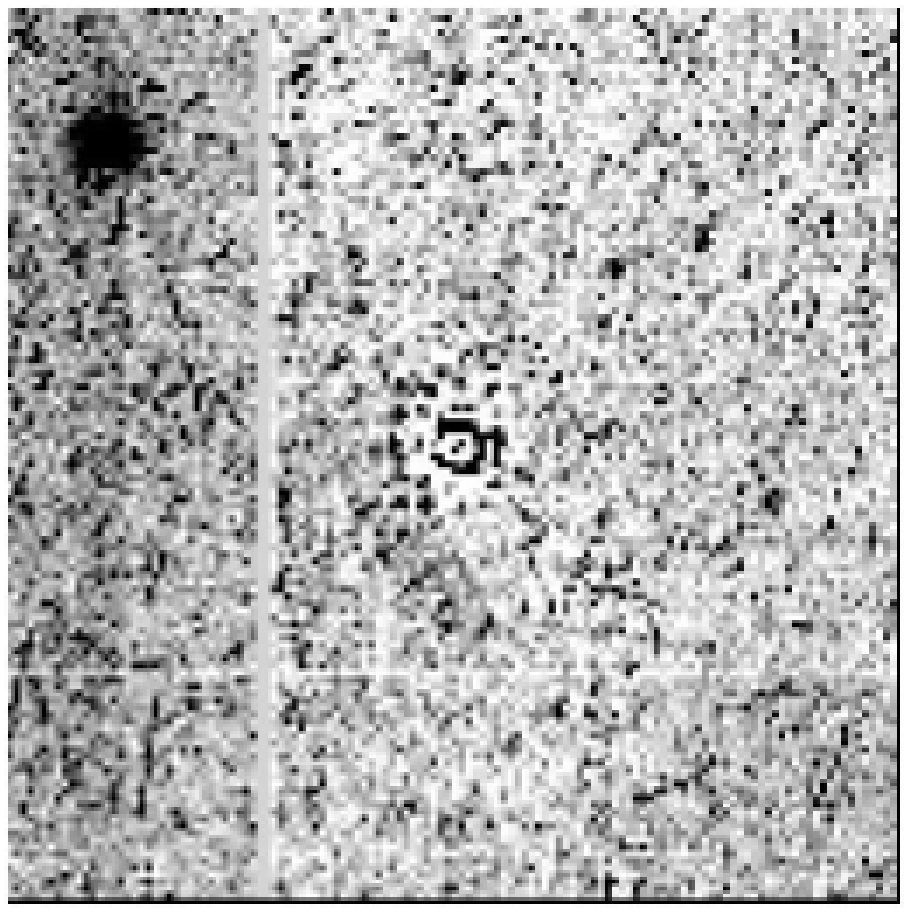} \\
   \includegraphics[width=4.4cm]{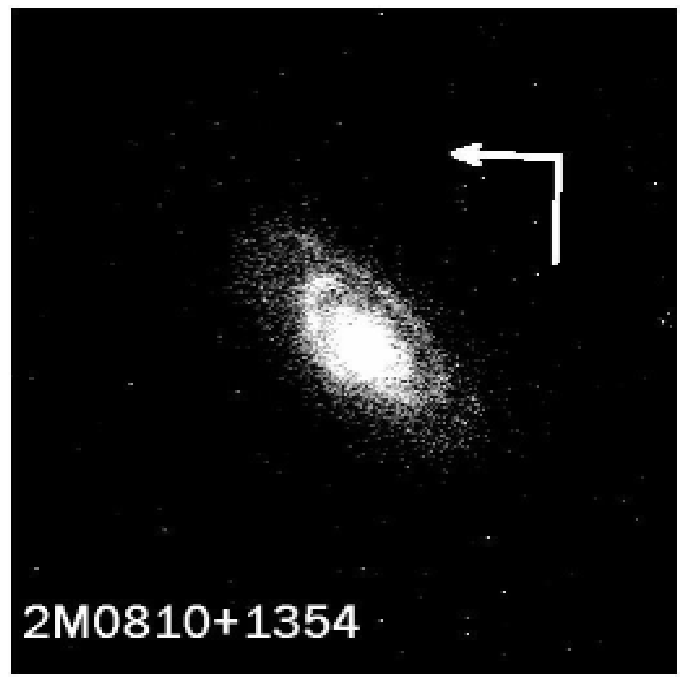}
   \includegraphics[width=4.4cm]{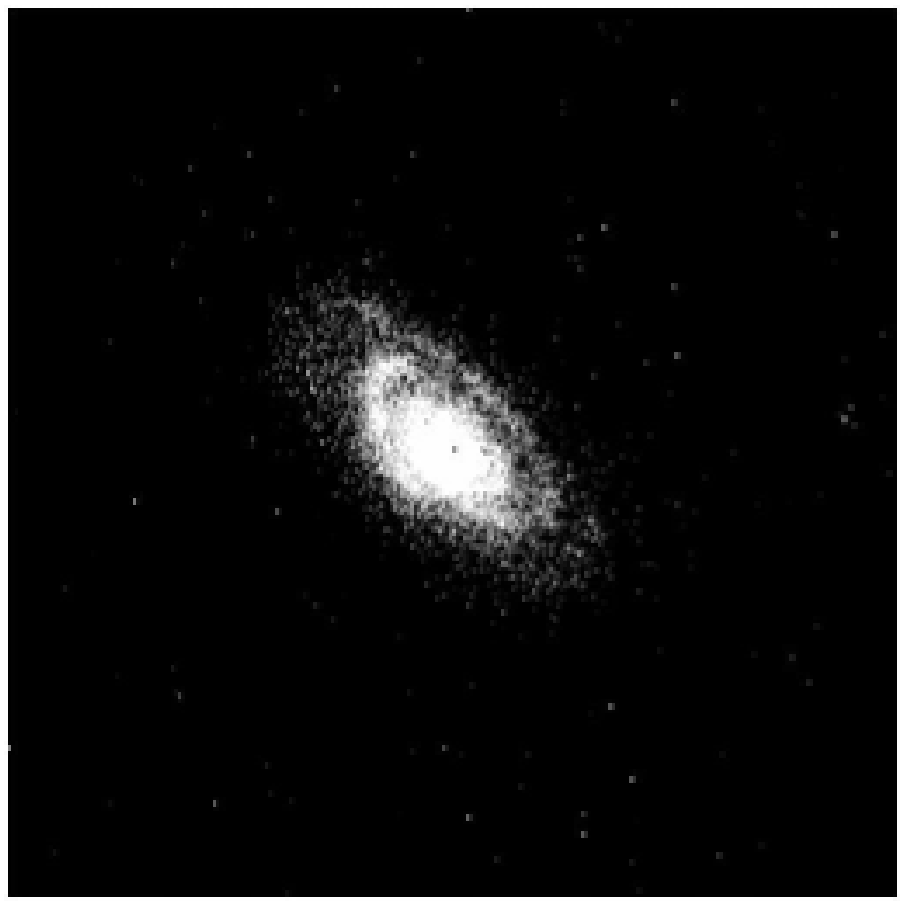}
   \includegraphics[width=4.4cm]{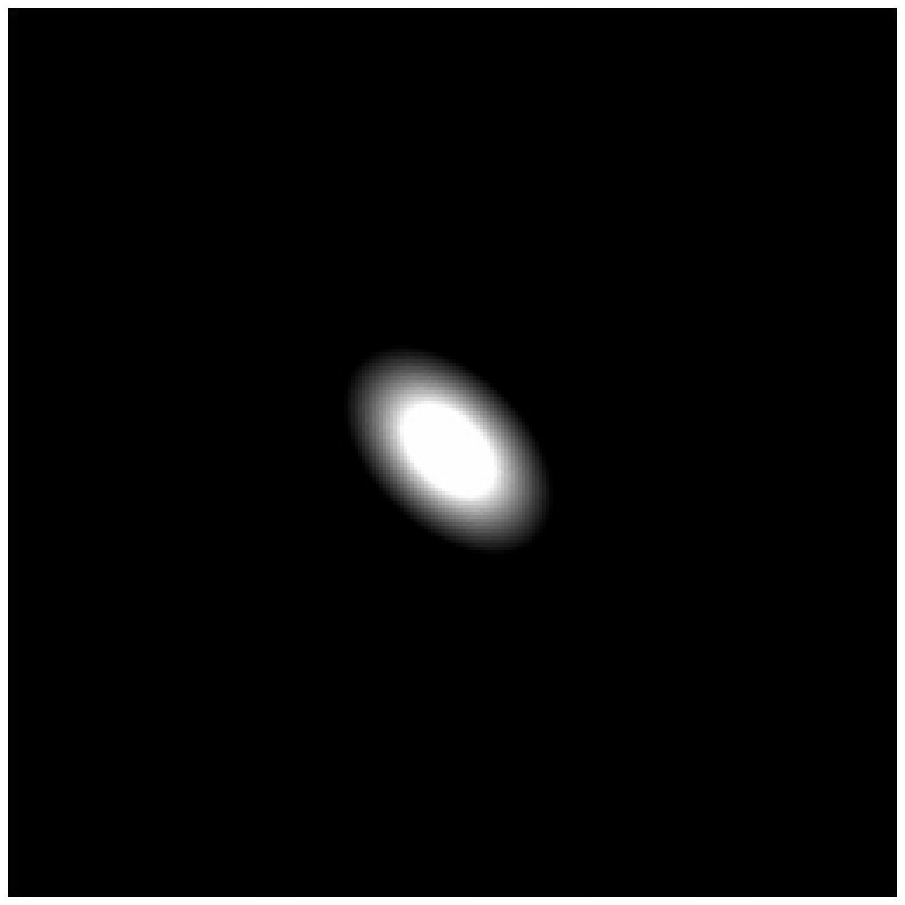}
   \includegraphics[width=4.4cm]{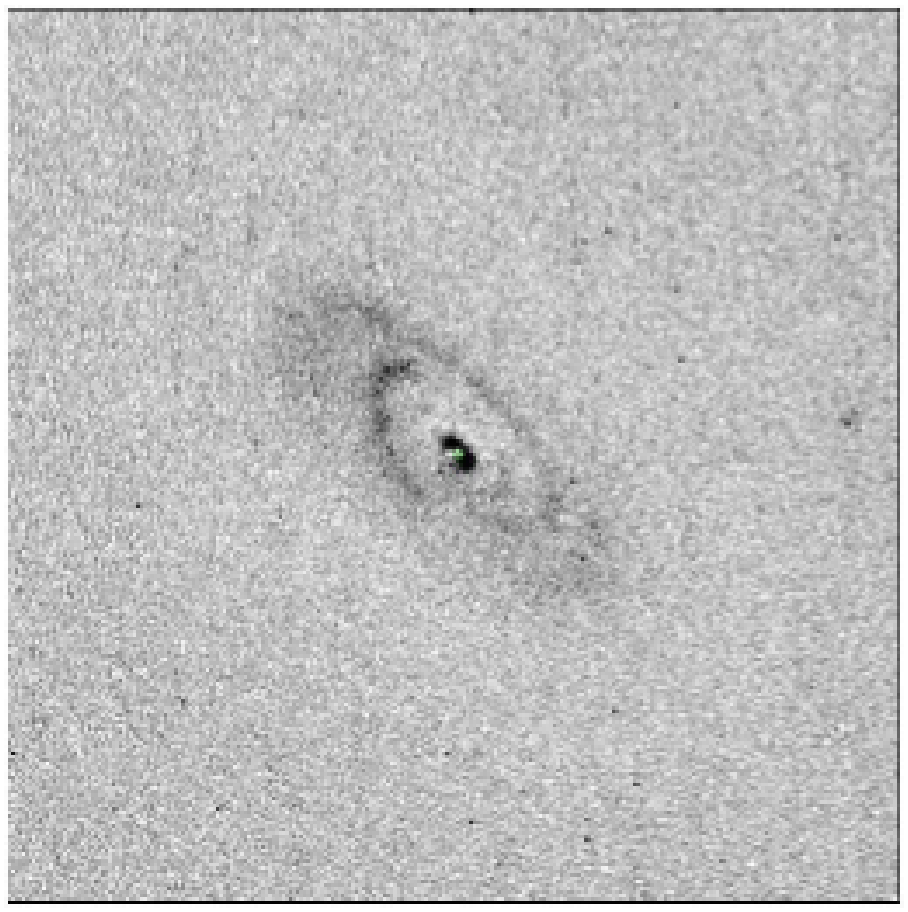}

       \caption{Typical image analysis process and results for
            two representative objects of our sample (see text for details). North and East are marked on the
            left most frames (with the arrow pointing to the North). The top series of images
            presents one of the RQQs observed in the near-IR domain by \emph{Ml01} :
            PG~0947+396. It shows that the PSF subtraction step is
            of prime importance to characterize the faint
            underlying host. The bottom series present the WFPC2
            imaged 2M~000810+1354 RQQ by \emph{Ma03}, hosted in a
            spiral galaxy.
	    }
         \label{resdec}
     \end{figure*}

    Fig.~\ref{resdec} illustrates the results for two
    representative
    objects of our \emph{new $PA_{host}$ data} sample. From left to right, we show for each
    object the final reduced HST image, a closer view on the PSF-subtracted
    image, the host galaxy model fitted over the PSF-subtracted
    image and the model-subtracted residual image (darker regions account for
    un-modelled details such as spiral arms, interaction features etc.).

\addtocounter{table}{1}

    \subsubsection{Morphology of the host}

    As we use only a \emph{de Vaucouleurs} model to derive the host galaxy
    parameters, we can not provide a host morphology
    classification on the basis of a better description of the
    host by a spheroidal or a disc-like profile.
    The morphological classification was thus realized by careful visual inspection of both the
    PSF-subtracted image and the model-subtracted residual image.
    We adopt a classification scheme similar to the one defined in
    Guyon et al. (\cite{gu06}) : if a bar or spiral arms are
    obvious in one image, the host is classified as a Spiral (disk-present) and abbreviated
    by S in Table~\ref{imres}. If no spiral structure or bar are visible in the
    image, we classify the host as an Elliptical galaxy (E). The remaining host galaxies (those presenting
    strong interaction traces/complex morphology, multiple nuclei, etc.) were assigned to the
    ``undefined" class (abbreviated U).
    Finally, in some cases no host galaxy was detected even after
    PSF-subtraction, essentially due to the comparatively
    high redshift or to the insensitivity to
    low surface brightness features due to short exposure times.

    \subsubsection{Reliability and error estimates}
    \label{compa}

      The $b/a$ and $PA_{host}$ reported in Table~\ref{imres} are
      determined within uncertainties that essentially come
      from the degeneracy existing between the host galaxy model and the
      estimated nuclear contribution. These uncertainties are quite difficult
      to estimate, especially in our case where the objects come
      from samples with different observational circumstances.

      A useful test to estimate the
      uncertainties of the derived parameters was proposed
      in \emph{Fl04} : to ensure that the modelling
      routines of the host reach the global minimum
      within the six-dimensional model parameter space, they
      used six different starting point for each object close to
      the best-fitting parameter value. Using the same method, we checked the quality of the
      determination of the host parameters by
      starting the MCS host modelling routine with initial
      parameters which progressively departs from the parameter set we
      found during the first modelling of each object. This enabled us
      to also check the stability of the solution found.

      This procedure allowed us to
      distinguish three cases that we relate to a quality
      criterion (Col.~(9) in Table~\ref{imres}). The first, called 1, accounts
      for situations where the modelling algorithm converges towards similar solutions
      whatever the starting point used, then giving reliable host parameters. The second, noted 2,
      corresponds to the cases where the solution strongly depends
      upon the initial parameters (generally corresponding to the
      faintest/rounder hosts). The last one, noted 3, is attributed to the cases where
      we detect no host galaxy after the PSF-subtraction step.

      Moreover, we have also considered among our sample of \emph{new $PA_{host}$ data} five objects which already had a $PA_{host}$ determined in the literature. We find a  good
     agreement\footnote{We find $\Delta PA_{host}$ within $10\degr$ for the class 1 quality objects, see Appendix~\ref{compapp} for more details.} between the previously
     published parameters and the one we derived,
     assessing the capability of the method used in our study to
     recover host galaxy parameters. This is not a surprise since
     the crucial step in the $PA_{host}$ determination essentially lies in the proper
     subtraction of the quasar nucleus, a step which is carefully carried
     out for each of the objects of our compilation (see Sect.~\ref{hostdet}).
     Finally we also carried out simple ellipse fitting on the nuclear-subtracted images
     and we observed that the derived $PA_{host}$ agree (within $10\degr$) with those derived
     using the 2D profile fitting procedure described in Sect.~\ref{decoprocess}. On the other hand
     the axial ratio $b/a$ is more sensitive to the PSF subtraction
     and then not reliable enough to be used in the study. Therefore, in the following, we only use $b/a$
     to cut the sample (cf. Sect.~\ref{statos}).

  \section{Polarimetric and radio data}
  \label{rapola}

\subsection{The polarimetric data}

    Most polarimetric data (the degree of polarization $P$ and the polarization
    angle $\theta_{Pola}$) are taken from a compilation of
    measurements from the literature (Hutsem\'ekers et al. \cite{hu05}). The polarimetric measurements
    for the objects of the sample of \emph{Za06} were published in their paper. For the
    2-MASS reddened objects of the \emph{Ma03} sample, the polarimetric data
    are taken from Smith et al. (\cite{sm02}).

    From this compilation we only considered and used data for objects where significant polarization
    is detected (i.e. such that $P/\sigma_{P} \geq 2$). This also ensures that we only select data for which the
    the polarization angle is known with an uncertainty of $\sigma_{\theta_{Pola}} \le 14\degr$.

    \begin{table}
\begin{minipage}[t]{\columnwidth}
\caption{Radio data for the Type~1 RLQs of
      our compilation which possess known $PA_{host}$ and $\theta_{Pola}$.}
\label{radiodata}
\centering
\renewcommand{\footnoterule}{}

      \begin{tabular}{llcrl}     
      \hline\hline

    Name & Catalog Name & $z$ & $PA_{Radio}$\footnote{$PA_{Radio}$ measured in degrees East of North.} & Ref.\footnote{References. (1) Bridle et al. \cite{bri94};
      (2) Price et al. \cite{pri93}; (3) Bogers et al. \cite{bo94}; (4) Reid et
      al. \cite{re95}; (5) Lister et al. \cite{li94}; (6) Hintzen et al. \cite{hi83};
      (7) Gower \& Hutchings \cite{go84a}; (8) Owen \& Pushell \cite{ow84}; (9)
      Gower \& Hutchings \cite{go84b}; (10) Readhead et al. \cite{re79}.} \\
      (B1950) & ~ & ~ & ($\degr$) & ~ \\ 

      \hline


0133+207    &   3C~47.0  &   0.425   &   35  &   \emph{Le99} \\
0137+012    &   4C~01.04 &   0.258   &   30  &   5,6,7   \\
0340+048    &   3C~393.0 &   0.357   &   40  &   \emph{Le99} \\
0518+165    &   3C~138.0 &   0.759   &   70  &   \emph{Le99} \\
0538+498    &   3C~147.0 &   0.545   &   60  &   \emph{Le99} \\
0710+118    &   3C~175.0 &   0.770   &   60  &   \emph{Le99} \\
0723+679    &   3C~179.0 &   0.846   &   90  &   \emph{Le99} \\
0740+380    &   3C~186.0 &   1.067   &   140 &   \emph{Le99} \\
0758+143    &   3C~190.0 &   1.195   &   30  &   \emph{Le99} \\
0906+430    &   3C~216.0 &   0.670   &   40  &   \emph{Le99} \\
1020-- 103    &  ... &   0.197   &   155 &   5,9    \\
1100+772    &   3C~249.1 &   0.311   &   100 &   \emph{Le99} \\
1111+408    &   3C~254.0 &   0.736   &   110 &   \emph{Le99} \\
1137+660    &   3C~263.0 &   0.646   &   110 &   \emph{Le99} \\
1150+497    &   4C~49.22 &   0.334   &   15  &   4,8 \\
1226+023    &   3C~273.0 &   0.158   &   60  &   10  \\
1250+568    &   3C~277.1 &   0.320   &   130 &   \emph{Le99} \\
1302-- 102    & ...  &   0.286   &   25  &   10  \\
1416+067    &   3C~298.0 &   1.436   &   90  &   \emph{Le99} \\
1458+718    &   3C~309.1 &   0.905   &   145 &   \emph{Le99} \\
1545+210    &   3C~323.1 &   0.266   &   20  &   3,9    \\
1618+177    &   3C~334.0 &   0.555   &   140 &   \emph{Le99} \\
1704+608    &   3C~351.0 &   0.371   &   40  &   1,2,4   \\
1828+487    &   3C~380.0 &   0.692   &   135 &   \emph{Le99} \\
2120+168    &   3C~432.0 &   1.785   &   135 &   \emph{Le99} \\
2135-- 147    & ...  &   0.200   &   100 &   9  \\
2247+140    &   4C~14.82 &   0.237   &   35  &   5   \\

      \hline
      \end{tabular}
      \end{minipage}
     \end{table}

\subsection{The radio data}
\label{radiogaga}

    For each of the objects from of our compilation which possess a $PA_{host}$
    and a $\theta_{Pola}$, we searched in the literature for radio
    data. The radio data, referring to the RLQs,
    consists of radio-jet position angles $PA_{Radio}$
    determined on the basis of published Very Large Array (VLA) maps from the
    literature.
    The position angle $PA_{Radio}$ refers to the orientation defined by the axis of the prominent jet (when seen),
    otherwise it is the position angle of the large scale radio
    morphology (brightest hotspot, outer lobes) relative to the core.
    As for the optical and near-IR $PA_{host}$, the angles
    are computed in degrees East of North.

    A large part of the Type~1 RLQ radio data comes from the
    measurements realized by \emph{Le99}.
    Other Type~1 RLQ radio-jet data were found in the literature for
    several quasars (Table~\ref{radiodata}).
    In the case of the Type~2 RLQs, the data come from the papers of \emph{Ci93}, \emph{Dk96} and \emph{Ma99}
    and essentially consist in compilations of VLA/Multi Element Radio Linked Interferometer Network (MERLIN)
    surveys measurements from the literature.

\section{Search for correlations}
 \label{statos}

    The Table~6 with all data is available on-line.
    For the analysis, we select the objects for which
    the data are relevant and accurate. Let us quickly summarize the
    criteria we use.

\begin{enumerate}

    \item{}We do not consider Seyfert galaxies in this study.
    To avoid these objects, we only consider objects with
    $M_{V} \la -23$ in the V\'eron catalogue
    version 12 (V\'eron-Cetty \& V\'eron \cite{ver06}). Two
    objects (0204+292 (\emph{Du03}, $M_{V}$= -22.4) and 2344+184 (\emph{Du03}, $M_{V}$= -22.0)) are then excluded from our
    sample. Two objects at the borderline (1635+119 (\emph{Du03},
    $M_{V}=$-22.8) and 0906+430 (\emph{Le99}, $M_{V}$=-22.9)) are
    kept. This $M_{V} \la -23$ criterion can not be crudely applied
    to the reddened objets of \emph{Ma03} and the Type~2 Radio-Quiet objects of
    \emph{Za06} since the luminosity of those objects is
    attenuated in the $V$-band. However, their high luminosity
    in the $K$-band or in the [\ion{O}{iii}] emission lines testifies their
    membership to the intrinsically luminous quasar sample.

    \item{} We decided to fix an axial ratio limit $b/a < 0.9$ in order to
    eliminate the roundest host galaxy morphology for which no accurate $PA_{host}$ determination
    could be obtained.

   \item{}When several data were available for a given object parameter, we
   always selected the value with the smallest uncertainty.

   \item{}For consistency, we always preferred data coming from
   larger homogeneous samples to those selected from smaller ones.

\end{enumerate}

 \begin{table}
 \begin{minipage}[t]{\columnwidth}
\caption{Results of the two sample Kolmogorov-Smirnov test applied to the
$\Delta \theta$ distribution of Type~1 and Type~2 quasars.}

\label{table:1}
\centering
\renewcommand{\footnoterule}{}

\begin{tabular}{lrrr}     
\hline\hline 

  Selected Sample\footnote{Criteria used to define sub-samples.} & $N_{1}$\footnote{Size of the selected Type~1 subsample.} & $N_{2}$\footnote{Size of the selected Type~2 subsample.} & $P_{K-S}$\footnote{Probability $P_{K-S}$ (in percent) that the $\Delta \theta$ of Type~1 and Type~2 objects are selected from the same parent sample.} \\
  ~ & ~ & ~ & (\%)\\

      \hline

 RQ                                    &  24   &   5  &  7.24  \\
 RQ ($P \geq 0.6 \%$)                  &  15   &   5  & 15.75  \\
 RQ ($z \geq 0.2$)                     &  8    &   5  &  0.11  \\
 RQ ($P \geq 0.6 \%~\&~z \geq 0.2$)    &  5    &   5  &  0.37  \\

\noalign{\smallskip}

 RL                                    &  16   &   16  &  2.30  \\
 RL ($P \geq 0.6 \%$)                  &  14   &   15  &  1.95  \\
 RL ($z \geq 0.2$)                     &  13   &   10  &  0.23  \\
 RL ($P \geq 0.6 \%~\&~z \geq 0.2$)    &  13   &   10  &  0.23  \\

\noalign{\smallskip}

 RQ+RL                                    &  40   &   21  &  0.24  \\
 RQ+RL ($P \geq 0.6 \%$)                  &  29   &   20  &  0.21  \\
 RQ+RL ($z \geq 0.2$)                     &  21   &   15  &  0.01  \\
 RQ+RL ($P \geq 0.6 \%~\&~z \geq 0.2$)    &  18   &   15  &  0.03  \\


\hline
\end{tabular}
\end{minipage}
\end{table}

%
   \begin{figure}
   \centering
   \includegraphics[height=6.0cm]{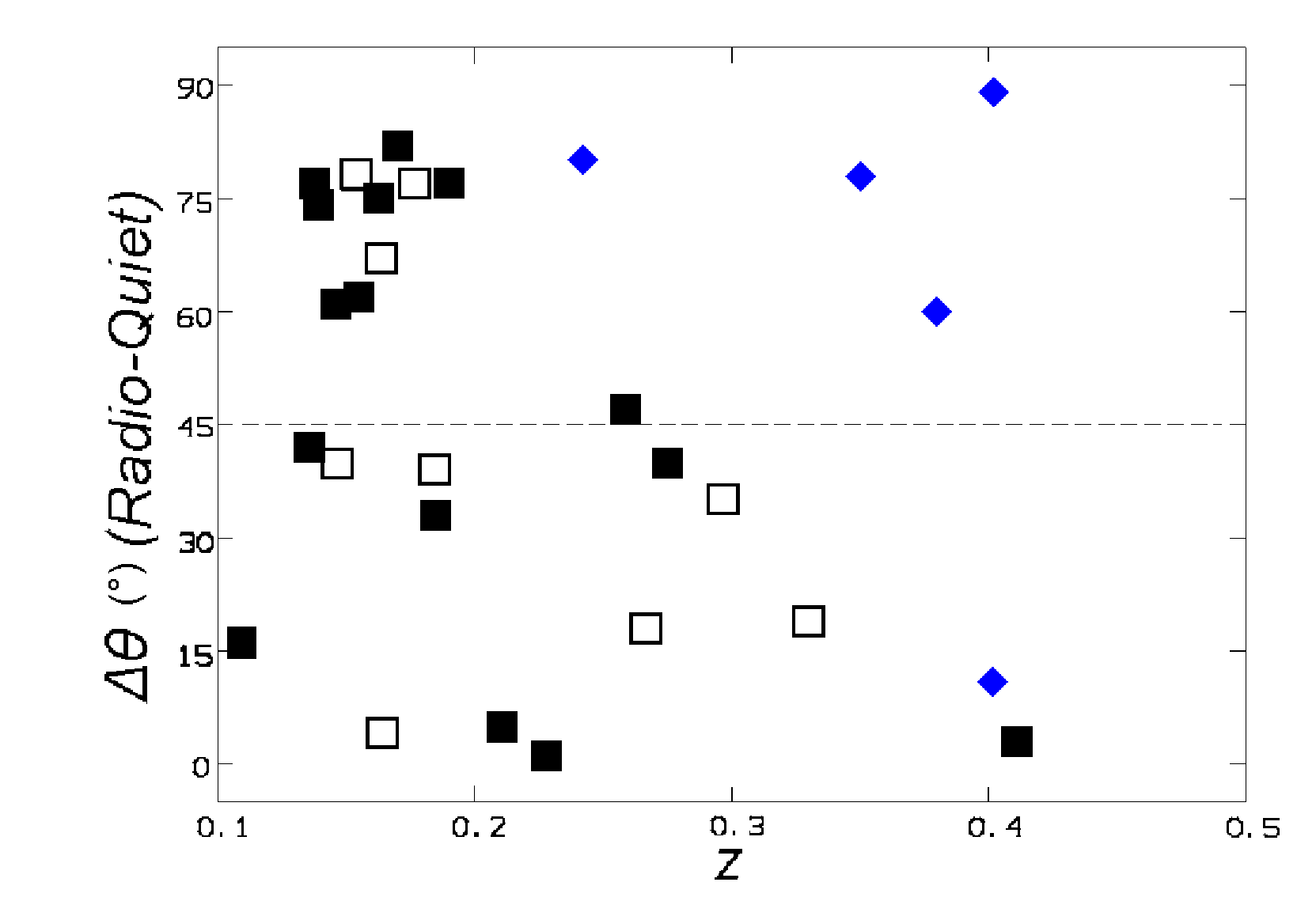}
   \\
  $ $
  $ $
   \includegraphics[height=6.0cm]{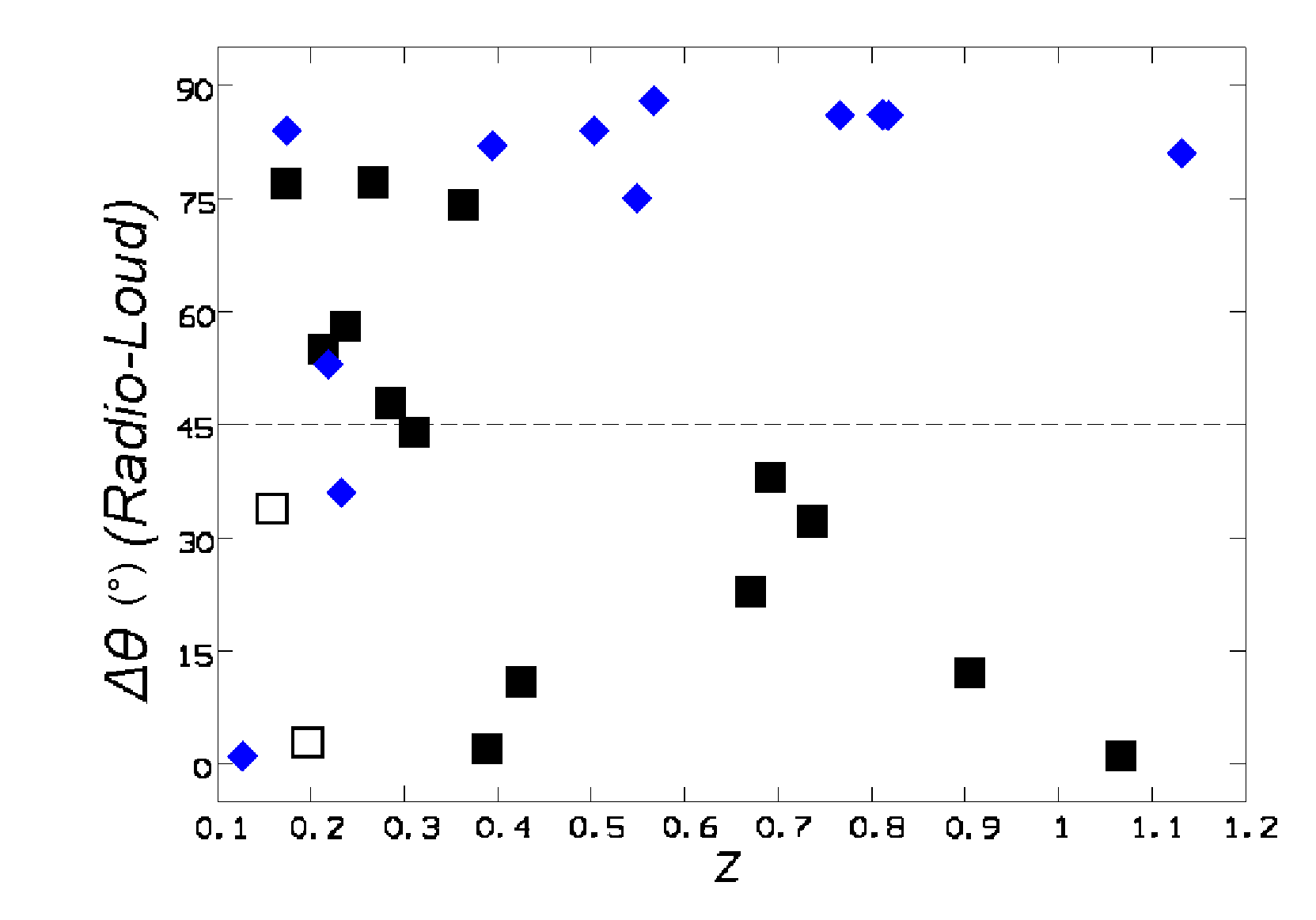}
   \caption{Distribution of the acute angle $\Delta \theta$ between the polarization position
            angle $\theta_{Pola}$ and the orientation of the host galaxy morphology $PA_{host}$,
            as a function of the redshift $z$. The upper panel refers to Radio-Quiet objects while the lower
            panel refers to Radio-Loud objects. Type~1 objects are
            represented in both panel by squares and Type~2
            objects by diamonds. The filled symbols refer to
            objects with a polarization degree $P \geq 0.6 \%$.}
              \label{rqdpa1z}%
    \end{figure}

   \subsection{The $PA_{host}$ - $\theta_{Pola}$ correlation}
       \label{opticor}

       Here we consider the host galaxy data that were derived
       using observations taken in the visible domain ($\lambda_{obs} \sim 6\,000$~\AA). The data
       obtained in the near-IR are discussed at the end of this section.
       We define $\Delta \theta$ as the acute angle between the polarization position angle $\theta_{Pola}$
       and the host galaxy orientation $PA_{host}$ in the visible : $\Delta \theta = 90-|90-|\theta_{Pola}-PA_{host}||$. Its value is
       defined between 0 and 90 degrees.

     \subsubsection{The Radio-quiet objects}

        The upper panel of Fig.~\ref{rqdpa1z} summarizes the $\Delta \theta$ behavior as a
        function of the redshift for both Type~1 and Type~2 RQQ objects in our sample.
        While at small redshifts there seems to be no
        clear correlation between $\theta_{Pola}$ and $PA_{host}$ for Type~1 objects
        (the redshift extent of the Type~2 RQQ sample is too limited to conclude),
        a distinct separation between the Type~1 and Type~2 behavior
        is noted for objects lying at larger redshifts ($z \ga
        0.2-0.3$) where the polarization position angle of Type~1 objects is preferentially parallel
        to the host major axis (``alignment" : $\Delta \theta \leq 45\degr$)
        while the polarization is mostly perpendicular to the host major axis in Type~2 quasars (``anti-alignment" : $\Delta \theta \geq 45\degr$).
        We also observe this type
        of behavior if we only consider the more polarized
        objects ($P \geq 0.6\%$ pictured by filled symbols).
        It is worth noting that the 2MASS reddened quasars from the \emph{Ma03} study appear preferentially
        at small $\Delta \theta$ values ($\Delta \theta \leq
        45\degr$) as other un-reddened Type~1 objects from the sample.
        We finally note that one Type~2 RQQ quasar at $z=0.402$ appears at
        small $\Delta \theta$. The behavior of this object, SDSS J0920+4531, already
        discussed in \emph{Za06}, seems to be due to an unfortunate apparent superposition of a
        close companion galaxy leading to a biased $PA_{host}$ determination.

        In order to assess the significance of the alignment
        observed in the upper panel of Fig.~\ref{rqdpa1z} we carried out statistical tests on the Type~1 and Type~2 RQQ samples. We used the two sample
        non-parametric Kolmogorov-Smirnov ($K$-$S$) test to
        determine if the Type~1 and Type~2 $\Delta \theta$ distributions are drawn from the same
        parent population. The results of the $K$-$S$ tests
        applied to the two aforementioned samples using several selection criteria are summarized in
        Table~\ref{table:1}. From these results, we can note that, while
        accounting for the whole RQQ sample does not allow
        an effective distinction between the Type~1 and Type~2 parent
        sample, the selection of the higher redshift objects
        ($z \geq 0.2$) shows that the probability that the
        difference in the $\Delta \theta$ behavior seen among the
        two samples is fortuitous is of only 0.11 \%. The same
        conclusions can be reached if we only consider the more
        polarized objects ($P \geq 0.6 \%$) but with a weaker
        statistical significance as the selected sample becomes smaller.

        We also investigated the variation of $\Delta \theta$ as a function
        of the redshift considering Type~1 and Type~2 objects separately.
        To this aim, we performed a non parametric \emph{Kendall-$\tau$} rank test
        on the Type~1 RQQ sample in order to check the strength of the correlation
        between the alignment ($\Delta \theta$) and the redshift.
        This statistical test uses the relative order of ranks in
        a data set to determine the likelihood of a correlation.
        We found a 4\% probability that the anti-correlation between
        the $\Delta \theta$ and the redshift
        is purely fortuitous in the case of the Type~1 RQQ sample.
        We do not carry this test on our Type~2 RQQ sample given its
        limited redshift range.

       \subsubsection{The Radio-loud objects}
         \label{refrl}

        The $\Delta \theta$ relation is also studied in the case of the Radio-Loud sample.
        The lower panel of Fig.~\ref{rqdpa1z} illustrates the observed $\Delta \theta$ behavior in this case.
        Once again, the $\Delta \theta$ distribution is clearly non-random.
        A first look allows to see that the Type~1 RLQs
        are preferentially found at small $\Delta \theta$ (i.e. alignment between the polarization angle
        and the major axis of the host). Type~2 objects often exhibit large
        offset angles, $\Delta \theta \geq 45\degr$, indicating an
    anti-alignment between the polarization angle and the
        host galaxy position angle (Cimatti et al. \cite{cima93}).

        The lower panel of Fig.~\ref{rqdpa1z} suggests a dependence of
        $\Delta \theta$ as a function of the redshift as already noted
        for the Type~2 RLQs (Cimatti et al. \cite{cima93}). Indeed, at small redshift ($z \le 0.2$), as
        in the Radio-Quiet case (upper panel of
        Fig.~\ref{rqdpa1z}), we do not find any particular $\Delta
        \theta$ difference for either Type~1 or Type~2 objects while
        at higher redshift ($z \ge 0.3-0.4$) we observe a clear
        separation of the $\Delta \theta$.

        Statistical tests were carried out on the data in
        order to assess the strength of the (anti-)alignment. The
        results of a two sample $K$-$S$ test using several selection criteria
        applied to Radio-Loud objects are summarized in Table~\ref{table:1}.
        The two sample $K$-$S$ test comparing the distribution of Type~1
        and Type~2 $\Delta \theta$ on the whole RLQ
        sample shows that there is only a 2.3\% probability that
        the $\Delta \theta$ of the two samples are selected from the same
        parent distribution. More selective
        criteria\footnote{Although the alignment/anti-alignment seems to appear at higher redshift
        for RLQs, we keep the cutoff in the test at $z \geq0.2$ for consistency with RQQs. However, using a higher
        redshift cutoff results in even more significant correlations.}
        ($z \ge 0.2$, $P \geq 0.6\%$ or both) on the RLQ sample reinforce the
        conclusion of the statistical test.

        A \emph{Kendall-$\tau$} correlation test
        considering Type~1 and Type~2 objects separately was applied to the sample in order to
        investigate the redshift dependency of the $\Delta
        \theta$ distribution. For the Type~1 RLQs, the result of
        the test shows that there is a 4\% probability that the
        anti-correlation of the $\Delta \theta$ with the redshift is purely fortuitous
        while for Type~2 objects the probability that the correlation is fortuitous is only of 0.2\%.

      \subsubsection{A global view}

      A global look at both panels of Fig.~\ref{rqdpa1z}
      suggests a similar behavior of the Type~1/Type~2 quasars whatever
      their radio-loudness: at $z \geq 0.2-0.3$ the Type~1 objects
      systematically lie at small $\Delta \theta$ having their polarization mainly parallel ($\Delta
      \theta \leq 45\degr$) to their morphological major axis while the Type~2
      objects are found at higher offset angle ($\Delta
      \theta \geq 45\degr$). These observations are supported
      by statistical tests carried out on the mixed RQQ
      and RLQ sample (Table~\ref{table:1}). The \emph{Kendall-$\tau$} correlation test lead to the
      conclusion that there is a 0.2 \% probability that  the correlation between $\Delta
      \theta$ and $z$ is purely fortuitous for the Type~2 quasars and of 0.4 \%
      that the anti-correlation is fortuitous for Type~1 objects.

      In order to verify that the difference in the behavior of $\Delta \theta$
      between Type~1 and Type~2 quasars actually corresponds to alignment and anti-alignment respectively,
      we computed the mean value $\overline{\Delta \theta}$ of the $\Delta \theta$ for both samples at $z \ge 0.2$
      separately and compared this value to the
      distribution of mean offset angles computed from simulated
      samples obtained by shuffling the $PA_{host}$ among the
      objects. We found that the $\overline{\Delta \theta} = 31\degr$
      (resp. $\overline{\Delta \theta} = 72\degr$) value measured for the Type~1 (resp. Type~2)
      quasars has a probability of 0.3 \% (resp. 0.02 \%) to occur in a random distribution
      of polarization and host position angles. This test was not
      applied to RQQ and RLQ samples separately given the
      small number of objects.

      We also investigated the $PA_{host}$ - $\theta_{Pola}$
      correlation for the RQQ and RLQ samples using the compilation of near-IR $PA_{host}$ data.
      However for those data, we do not see any clear correlations between the polarization
      angle and the host galaxy position angle even at higher redshifts ($z \ge 0.2-0.3$).
      This lack of correlation was confirmed by the $K$-$S$
      and \emph{Kendall-$\tau$} tests.

   \subsection{Other interesting correlations}
     \label{othercorr}

   \begin{figure}
   \centering
   \includegraphics[height=9.0cm,angle=270]{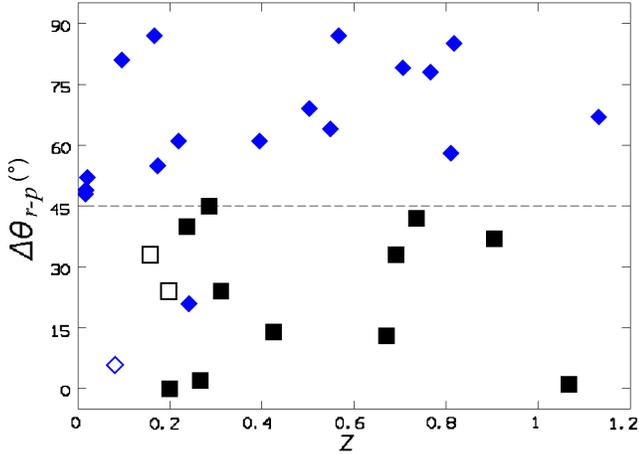}
   \caption{Distribution of the acute angle $\Delta \theta_{r-p}$ between the
   polarization position angle $\theta_{Pola}$ and the
   orientation of the radio jet $PA_{Radio}$ defined in Sect.~\ref{radiogaga}. As in the other
   figures, Type~1 objects are represented by squares and Type~2 objects
   are represented by diamonds. The symbols are filled when $P \ge 0.6 \%$.}
              \label{rldparad}%
    \end{figure}

   \subsubsection{The $PA_{Radio}$ - $\theta_{Pola}$ correlation}
      \label{resradstat}

       Using the available VLA radio-jet orientation
       for the Radio-Loud objects of our sample
       possessing optical polarization measurements (Table~\ref{radiodata} and \emph{Ci93}), we
       also investigate the $PA_{Radio}$ - $\theta_{Pola}$
       correlation. We define the radio-jet/polarization offset angle in
       the same way as the offset defined in
       Sect.~\ref{opticor}: $\Delta \theta_{r-p} = 90-|90-|\theta_{Pola}-PA_{Radio}||$ whose value is
       comprised between 0 and 90 degrees.

       Fig.~\ref{rldparad} synthesizes the
       $\Delta \theta_{r-p}$ distribution for the Radio-Loud
       objects as a function of the redshift. One can note that the radio-jet of Type~1 Radio-Loud quasars are
       mostly aligned with the polarization ($\Delta \theta_{r-p} \leq
       45\degr$) while these two directions are mostly anti-aligned in Type~2 objects ($\Delta \theta_{r-p} \geq
       45\degr$). Moreover, we do not see any redshift
       dependence of the correlation. We carry out $K$-$S$ test
       over the $\Delta \theta_{r-p}$ distribution of Type~1 and
       Type~2 objects. The two sample $K$-$S$ test shows that there
       is less than a 0.001 \% probability that both samples
       are selected from the same parent distribution. This result
       is even more significant if we only consider the $P \geq 0.6$ \% quasars.
       We also find a 0.17 \% (resp. 0.06 \%)
       probability that the $\overline{\Delta \theta}_{r-p} = 24\degr$ measured
       for Type~1 RLQs (resp. $\overline{\Delta \theta}_{r-p} = 65\degr$ for Type~2) comes
       from a random $\Delta \theta_{r-p}$ distribution.

       These (anti-)alignment effects were already reported
       previously in the literature. For the Type~1 objects,
       a clearly non-random distribution of $\Delta \theta_{r-p}$ has been reported by
       many authors (e.g. Stockman et al. \cite{sto79}; Rusk \& Seaquist \cite{rusk85};  Berriman et al. \cite{beri90};
       Lister \& Smith \cite{li00}),
       showing that the optical polarization angle of Type~1 RLQs mostly
       aligns with their radio-morphology.
       For the Type~2 Radio-Loud quasars the anti-alignment between the radio-jets
       and the optical polarization is known as part of the so called ``\emph{RG alignment
       effect}" observed in high-$z$ radio galaxies (e.g. McCarthy et al. (\cite{ma87});
       Chambers et al. (\cite{cha87}); Cimatti et al. (\cite{cima93})). Indeed, using polarization
       measurements and VLA radio observations of RGs,
       they found a strong anti-alignment between the
       linear polarization orientation and the radio structure.

        \begin{figure}
   \centering
   \includegraphics[height=9.0cm,angle=270]{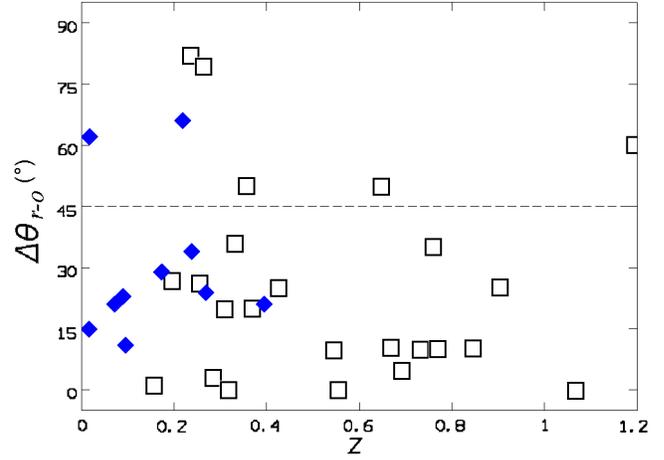}
   \caption{Distribution of the acute angle $\Delta \theta_{r-o}$ between the
   orientation of the radio jet $PA_{Radio}$ and the host galaxy position angle $PA_{host}$ in the visible domain.
   Type~1 objects are represented by empty squares and Type~2 objects
   are represented by filled diamonds.}
              \label{rldparadopt}%
    \end{figure}

   \subsubsection{The $PA_{Radio}$ - $PA_{host}$ correlation}

       Using the radio-jet orientations given in
       Table~\ref{radiodata} and \emph{Ci93} for the Radio-Loud objects of our sample with a
       reliable host galaxy position angle, we finally look at
       the correlation of the radio-jet with the morphology of the
       host. We define the offset angle $\Delta \theta_{r-o}$
       between the radio-jet and host major axis position angle
       using the same definition as in the preceding sections.
       Fig.~\ref{rldparadopt} summarizes the behavior of $\Delta \theta_{r-o}$
       as a function of the redshift.

        For the Type~1 RLQs, we note a significant alignment
       ($P \le 0.001 \%$ that the measured $\overline{\Delta \theta}_{r-o} = 24\degr$ is due to chance)
       between the host galaxy's main orientation and the
       radio-jet orientation, as first reported within a larger sample in \emph{Le99}.
       This is not a surprise since $2/3$ of the Type~1 RLQ sample used to check this correlation
       is made of the measurements given in \emph{Le99}.

       For the Type~2 RLQs (the NLRGs) we also note an alignment ($\Delta \theta_{r-o} \leq 45\degr$) between
       the radio-jet and the host position angle. The permutation test gives a $P \le 8\%$ that the measured $\overline{\Delta \theta}_{r-o} = 30\degr$ is due to chance. Although one might suspect a redshift dependence on
       Fig.~\ref{rldparadopt}, it is not significant in our sample (as the result of the \emph{Kendall-$\tau$} test). Let us notice that this alignment is known as part of the ``\emph{RG alignment effect}" (e.g. McCarthy et al. (\cite{ma87});
       Chambers et al. (\cite{cha87})) where the optical morphology of higher redshift ($z \ge 0.5$) RGs
       aligns with the radio-jet.

\begin{table}
\begin{minipage}[t]{\columnwidth}
\caption{Summary of the correlations studied.}

\centering
\renewcommand{\footnoterule}{}
\label{recapknown}

\begin{tabular}{clll}     
\hline \hline
    Radio Type & Correlation & Type~1\footnote{Numbers in parentheses give the reference to the
first detection of a given correlation for Type~1 and Type~2 quasars : $(1)$ Chambers et al.
\cite{cha87}; McCarthy et al. \cite{ma87}; Cimatti et al.
\cite{cima93}; $(2)$ \emph{Le99}; (3) Stockman et al.
\cite{sto79}; (4) Biretta et al. \cite{bir02}; (5) Zakamska et al.
\cite{za06} and (6) this work.} & Type~2$^a$\\

\hline 

    Radio-Quiet & $PA_{host} - \theta_{Pola}$         & $\parallel^{\mathrm{(6)}}$ & $\perp^{\mathrm{(5)}}$ \\
    Radio-Loud & $PA_{host} - \theta_{Pola}$          & $\parallel^{\mathrm{(6)}}$ & $\perp^{\mathrm{(1)}}$ \\
    Radio-Loud & $\theta_{Pola} - PA_{Radio}$         & $\parallel^{\mathrm{(3)}}$ & $\perp^{\mathrm{(1)}}$\\
    Radio-Loud & $PA_{host} - PA_{Radio}$             & $\parallel^{\mathrm{(2,6)}}$ & $\parallel^{\mathrm{(1,4)}}$\\

\hline
\end{tabular}
\end{minipage}
\end{table}

\subsection{Summary of the results}
   \label{summer}

  The known correlations between quasar orientation parameters
  are summarized in Table~\ref{recapknown}.
  The main result of our study is the discovery of a
  correlation between $PA_{host}$ and $\theta_{Pola}$ in quasars observed whatever their radio-loudness
  but depending on the spectroscopic type of the source.
  We find that while high redshift Type~2 quasars exhibit an
  anti-alignment of their optical polarization angle with their optical, rest frame
  UV/blue, morphology, the optical host
  galaxy position angle is preferentially aligned with the polarization angle in Type~1 quasars.

  Another interesting result is the redshift
  dependence of the alignment. Indeed, as suggested by both upper and lower panels of Fig.~\ref{rqdpa1z}
  and the results of the statistical tests, the alignment becomes significant if we consider redshifts
  higher than $z \sim 0.2-0.3$ for either Radio-Loud or Radio-Quiet
  objects. This observation may explain the lack of correlation
  observed by Berriman et al. (\cite{beri90}) between
  $\theta_{Pola}$ and $PA_{host}$ since the redshift extent of
  their sample was rather small (among their 24 objects with $z \le 0.4$, only 5 of them
  have $z \ge 0.3$). Note that the precise cutoff is not clearly
  determined and might be slightly higher as suggested by RLQs
  (Fig.~\ref{rqdpa1z}). Observations of RQQs at higher $z$ would
  be needed to clarify this point.

  Finally, we note that, while the alignment between the host galaxy position angle and the
  polarization angle is clearly observed and statistically significant in the case of the optical
  $PA_{host}$ data, we do not find a similar behavior with the near-IR $PA_{host}$ data.
  Optical and K-band studies of high $z$ RGs already showed that while an alignment effect
  is observed between their optical emission and their radio axis, a significantly weaker
  alignment is observed between the K-band emission and the radio axis (Rigler et al. \cite{ri92};
  Dunlop \& Peacock \cite{du93}; Best et al. \cite{be98}). This argues in favor of a scheme where the
  K-band emission is not necessarily related to the visible (rest-frame blue) light.

\section{Discussion}
\label{discu}

 \subsection{The observed correlation}

 \begin{figure}
   \centering
   \includegraphics[height=6.0cm]{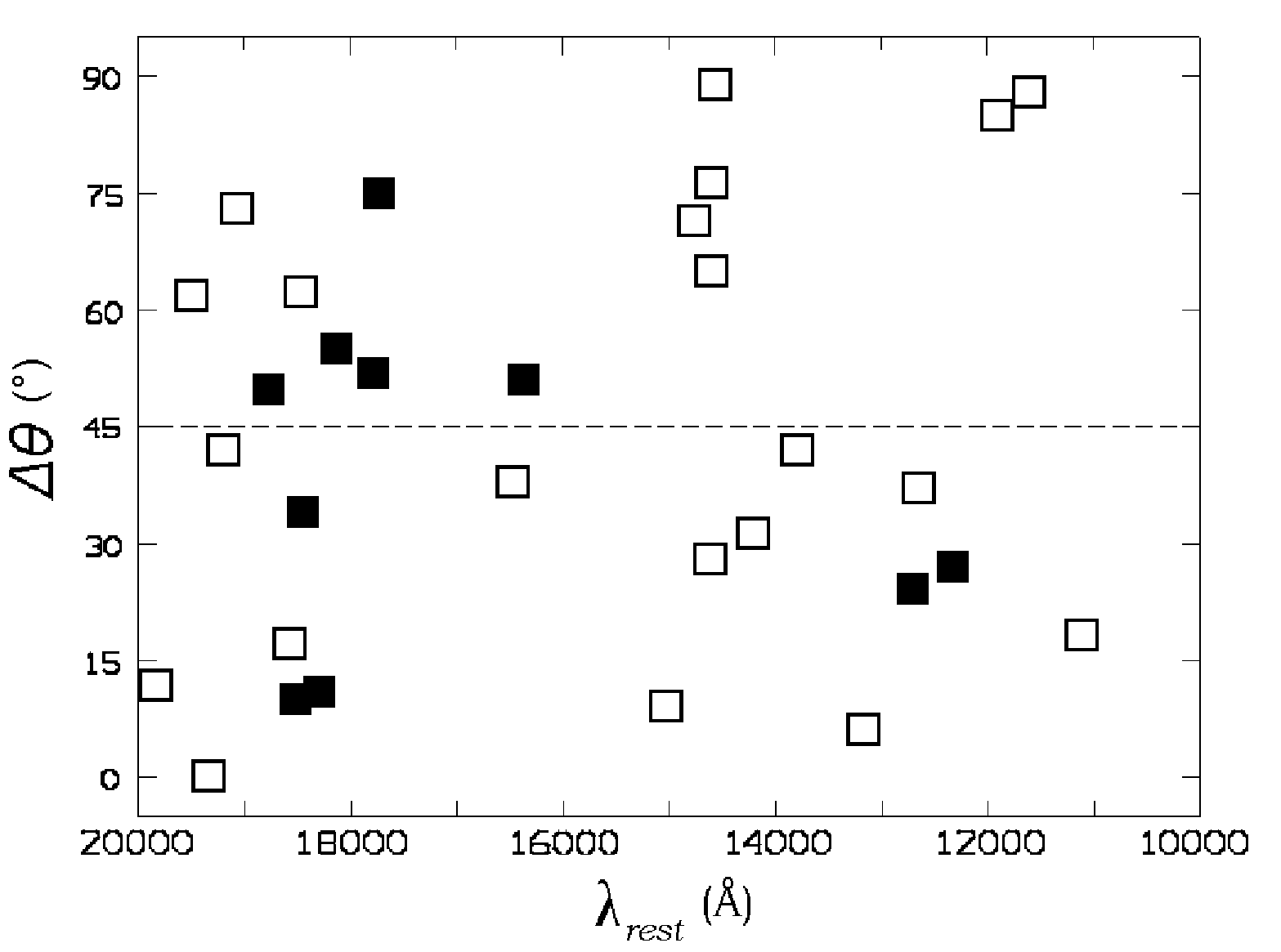}
    \\
  $ $
  $ $
  \centering
   \includegraphics[height=6.0cm]{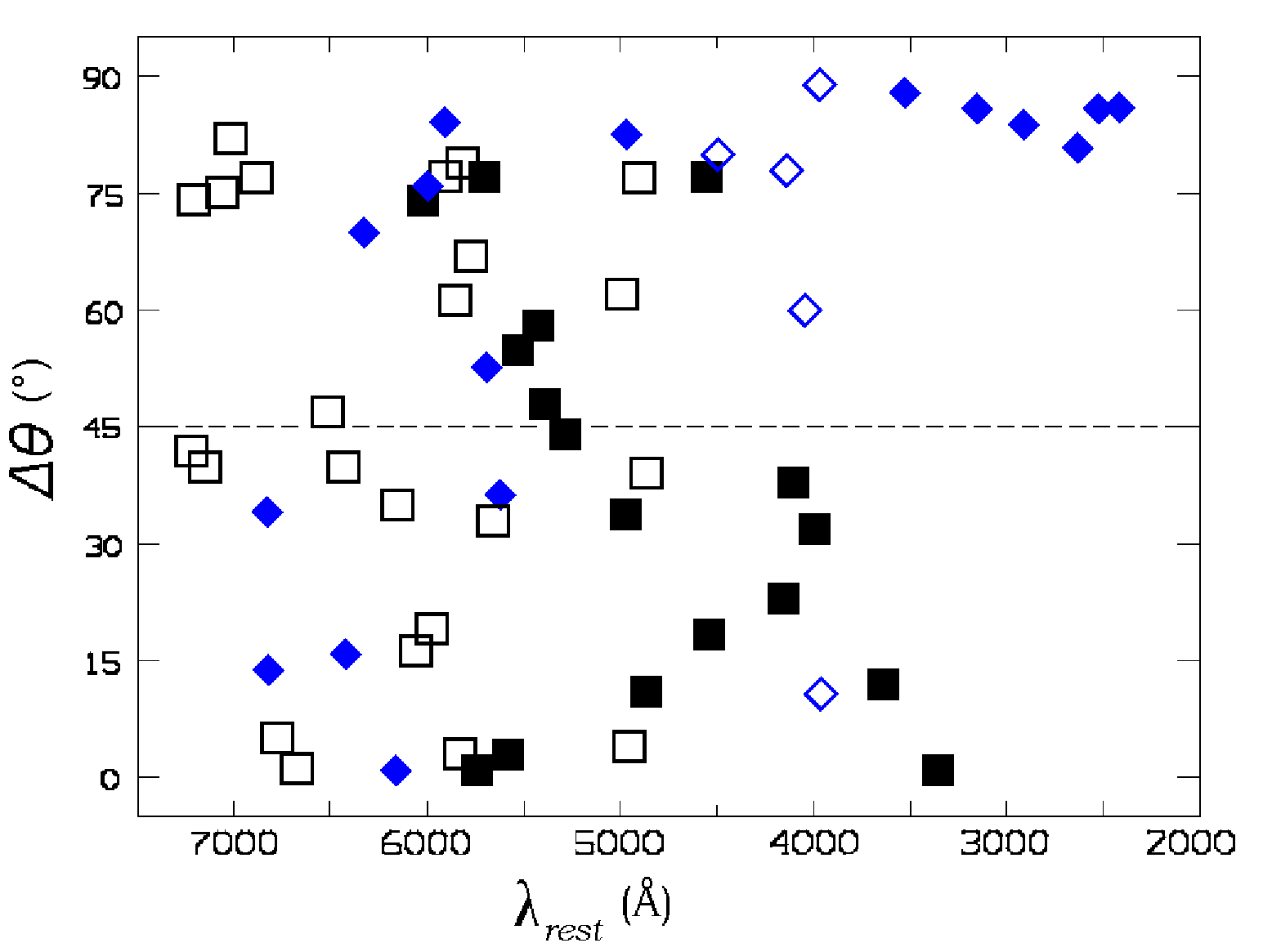}
   \caption{Distribution for RQQs and RLQs of the acute angle $\Delta \theta$ between $PA_{host}$ and $\theta_{Pola}$ as a function of the observation wavelength $\lambda_{rest}$ as measured in the quasar rest-frame. The upper panel presents the $\Delta \theta$ distribution at near-IR wavelengths while the lower panel presents the same distribution at UV/visible wavelengths. RQQs are represented by empty symbols and RLQs by filled symbols. Type~1 objects are pictured by squares while Type~2 objects are pictured by diamonds. No observations of Type~2 quasars are available in the near-IR. Our sample does not contain quasars imaged at $7500$~\AA~ $\le \lambda_{rest} \le$ $11000$~\AA.}

              \label{lambada}%
    \end{figure}

    In Sect.~\ref{statos} we reported
    on the existence of an (anti-)alignment between the
    linear polarization and the host galaxy position
    angle of (Type~2) Type~1 quasars. We noticed that this
    behavior seems to be independent of the radio-loudness. It is also interesting to remind the lack of
    correlation in the near-IR domain and the redshift dependence
    of the $PA_{host}-\theta_{Pola}$ (anti-)alignment.
    These last two facts lead us to think that the observed correlations
    might be linked to the rest-frame UV/blue part of the quasar light which is
    redshifted to optical wavelengths for the higher $z$ objects.
    This hypothesis is supported by Fig.~\ref{lambada}, where we plot $\Delta \theta$
    as a function of $\lambda_{rest}$,~the rest-frame wavelength at which $PA_{host}$
    is measured (i.e. the central wavelength of the imaging filter divided by $1+z$).

    We note that while at longer wavelengths there is no particular behavior in the
    $\Delta \theta$ distribution for either Type~1 or Type~2
    objects, the (anti-)alignment
    effect clearly appears at shorter wavelengths ($\lambda_{rest} \la 5000$~\AA).
    At $\lambda_{rest} \le 5000$ \AA, a two sample $K$-$S$ test shows that there is a 0.01 \%
    probability that the $\Delta \theta$ distribution of Type~1 and Type~2 quasars is drawn from the same parent sample.

\subsection{The UV/blue continuum and the origin of the correlation}

    The morphology of Type~2 Radio-Loud objects (NLRGs) has been extensively
    studied in the literature since those objects do not possess a bright central nucleus masking the
    underlying host and can be observed at
    high redshifts. Imaged at optical wavelengths, higher redshift RGs ($z \ge 0.5$) are known to
    show the aforementioned ``\emph{RG alignment effect}" where
    the extended rest frame UV/blue continuum emission aligns
    with the radio-jet. This alignment, not seen in the near-IR
    images of the same objects (Rigler et al.
    \cite{ri92}) leads to a two
    component RG model: at shorter wavelengths, the continuum is
    dominated by an extended UV/blue component aligned with the
    radio axis while at longer wavelengths, the quasar light is
    dominated by a stellar component which possesses the morphology and
    physical dimensions of elliptical galaxies.

     This extended UV/blue continuum which can show a quite complex morphology (McCarthy et al. \cite{ma97})
     was primarily interpreted, given its alignment with the radio-jet, as massive star formation regions triggered
     by the shocks associated to the passage of the radio-jet through the host galaxy (McCarthy et al. \cite{ma87}).
     However, as we briefly recalled in Sect. \ref{refrl}, high
     redshift NLRGs are also known to show an anti-alignment
     between their optical linear polarization and their rest frame
     UV/blue morphological structure (e.g. Cimatti et al.
     \cite{cima93}; Hurt et al. \cite{hu99}), supporting the possibility that at
     least part of this extended UV/blue light is due to the
     polar scattering of the light
     from the quasar nucleus by electrons and/or dust (di Serego Alighieri et al. \cite{di89}; Cimatti et al. \cite{cima93};
     di Serego Alighieri et al. \cite{di93};
     Dey et al. \cite{dey96};  Tadhunter et al. \cite{tad02}).
     Indeed, this assumption provides a simple explanation to the
     fact that the polarization angle is perpendicular to the
     optical, extended rest frame UV/blue light in the host galaxy.

     Due to their apparent faintness and the absence of strong
     radio counterpart, Type~2 Radio-Quiet quasars have
     long been searched for. The hard X-ray and SDSS
     surveys recently unveiled large samples of Type~2 RQQ candidates,
     showing that these objects are not extraordinarily rare (e.g. Zakamska et al. \cite{za03}
     and references therein). The HST spectropolarimetric study
     of a sample of 12 Type~2 RQQs revealed that a large part of them (9 out of 12) possess
     high optical polarization and that for 5 objects their polarized spectra
     contain broad emission lines betraying the presence of a Type~1 nucleus (Zakamska et al. \cite{za05}).
     Three band HST imaging of 9 Type~2 RQQs also reveals some morphologically complex
     emission which appears in the bluer images (\emph{Za06}). These structures
     were interpreted as scattering cones since their position angle lies almost
     perpendicularly to the $E$-$vector$ polarization plane. As in NLRGs, scattering
     in the extended UV/blue component seems to
     represent a viable interpretation of the origin of the
     detected polarization.

     The detection of a hypothetical extended blue component in
     either Radio-Loud or Radio-Quiet Type~1 quasars is hindered by the central source
     whose strong contribution at smaller wavelengths remains difficult to estimate and properly subtract.
     However, if Type~1 and Type~2 objects are, according to the
     unification model, intrinsically identical, the extended scattering
     region resolved in Type~2 quasars should also be present,
     but masked by the higher apparent brightness of the unobscured
     central source. Since the viewing angle of Type~1 and
     Type~2 objects is thought to be an essential parameter
     controlling the observed properties of those objects, the
     polarization geometry may also be affected by the viewing angle. While
     in this framework Type~2 objects are almost seen edge-on such that the extended polar scattering
     regions produce a polarization angle perpendicular to
     the symmetry axis of the AGN (pictured by the radio-jet axis
     in the RLQ objects), the situation may be different in the nearly face-on Type~1
     quasars.

\begin{figure}
   \centering
   \includegraphics[width=8.5cm]{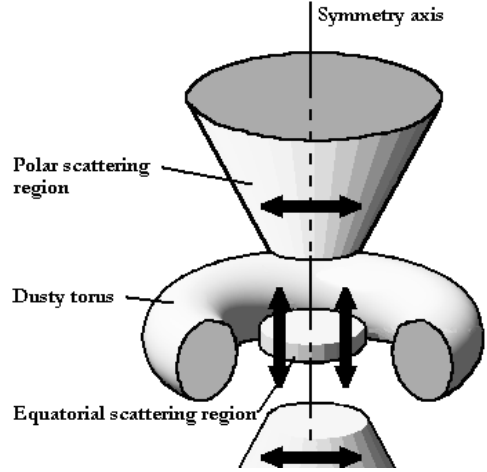}
   \caption{Cartoon illustrating the adopted quasar model and more
       particularly the two scattering regions used to interpret our results
(adapted from Smith et al.~\cite{sm04}). The system is shown inclined with respect
to the line of sight to the observer such that a Type 1 object is
seen (in Type 2 objects the dusty torus is seen edge-on and
the central part is hidden). The arrows indicate the direction of the resulting polarization projected onto the plane of the sky for each scattering region (see text for more details).
In RLQs, the radio jet (not illustrated)
extends along the symmetry axis.
   }
              \label{twoc}%
    \end{figure}

     An enticing interpretation of the polarization properties was recently proposed for
     the Seyfert galaxies. Indeed, Seyfert galaxies exhibit some kind of
     \emph{alignment effect} between their optical linear polarization
     and their radio axis. While the Type~2 objects polarization
     is known to be anti-aligned with the radio axis (Antonucci \cite{anto83}),
     the polarization originating from scattering cones
     perpendicular to the obscuring torus (Antonucci \& Miller \cite{anto85}), Type~1 Seyferts are
     found to be of both kinds (aligned and anti-aligned, Smith et al. \cite{smi02}). The
     scenario proposed to account for these observations consists of a
     two component polar+equatorial scattering model (Smith et al. \cite{sm04},
     \cite{sm05}; Goosmann et al. \cite{goo07}). While the extended
     polar scattering region accounts for the perpendicular polarization, the much smaller equatorial
     disk-like component inside the obscuring torus produces a polarization which, projected
     onto the plane of the sky, is parallel to the symmetry axis of the system (see Fig.~\ref{twoc}).
     In this framework, Type~2 polarization properties are dominated by polar scattering,
     the equatorial component being hidden by the dusty torus. In Type~1 objects, the
     smaller but significantly non-zero inclination angle of the system is such that both scattering
     regions contribute to the polarized flux, the resulting polarization being dominated
     by the equatorial scattering component (Smith et al. \cite{smi02}).

     This two component scattering scheme seems to nicely fit our observations as
     it provides an explanation to the correlations found in
     Sect.~\ref{opticor} assuming that the observed polarization is
     due to scattering (in both RQ and RL objects) and that the morphology of
     the host galaxy/extended emission observed in higher
     redshift quasars is, at least in part, related to scattered UV/blue light\footnote{Let us emphasize that
     only the polar scattering region --assumed to be parallel to the symmetry axis of the system-- can be angularly resolved (in principle in both Type~1 and Type~2 objects). The equatorial scattering region -perpendicular to the symmetry axis- is too small to be resolved.
      Note that in this framework, given the projections effects and the presence of two competing
      scattering regions, we expect the dispersion of the $\Delta \theta$ to be higher for Type~1 quasars
      than for Type~2 ones as it might be suspected in Fig.~\ref{rqdpa1z}.}.
     The model also provides a simple explanation of the alignment seen in Type~1 and Type~2 Radio-Loud
     quasars between their polarization position angle and their
     radio jets since the radio-jets are thought to be emitted along
     the symmetry axis of the accretion disk which also defines the direction of the extension of the polar scattering regions. Moreover, assuming the proposed model applies and reminding the lack of correlation in the near-IR (which samples the stellar component), our observations suggest that the central engine symmetry axis and the orientation of the stellar component are not correlated. This lack of correlation has already been reported in the case of Seyfert and radio galaxies (e.g. Kinney et al. \cite{ki00}~and references therein; Schmitt et al. \cite{sch01}) suggesting alternative models to the fuelling of the AGN supermassive black hole (e.g. King \& Pringle \cite{ki07}). Our results suggest that such scenarii might also apply to the higher luminosity quasars.
     
     Recent HST two band imaging of Type~1 RQQs (e.g. Sanchez et al. \cite{san04}; Jahnke et al. \cite{ja04})
     shows an enhancement of the rest frame blue
     light of the host galaxy with respect to normal galaxies of
     similar redshift and luminosity. This blue excess is actually
     explained as due to merger induced activity. However, those
     objects do not show more evidence for interaction than inactive
     galaxies at similar redshift and luminosity (Sanchez et al. \cite{san04}).
     Polarimetric study of these objects might provide an
     explanation of this excess blue light in terms of scattering, although an accurate estimate
     of the fraction of scattered light in Type~1 quasars might be difficult
     given the fact that the observed polarization is the combination of both polar and
     equatorial polarization acting in opposite ways.

     Let us finally note that we also searched --without success-- for a possible correlation between
     $\Delta \theta$ and the quasar absolute magnitude, since the fraction of scattered over stellar light may be
     more important in the hosts of intrinsically more luminous quasars.

\section{Conclusions}
\label{conclu}

    Using host galaxy position angles ($PA_{host}$) determined from high resolution optical/near-IR images of quasars
    and data from the literature, we investigate the possible existence
    of a correlation between the host morphology and the polarization direction in the case
    of Radio-Quiet and Radio-Loud quasars.
    We can summarize our results as follows :

   \begin{enumerate}

      \item We find an alignment between the direction of the linear polarization
      and the rest-frame UV/blue major axis of the host galaxy of Type~1
      quasars. In the case of Type~2 objects, it is well established that the extended UV/blue
      light is correlated to the observed polarization, as
      these blue regions are thought to be dominated by scattering. Our results suggest that such an extended UV/blue scattering region is also present in Type~1 quasars.

      \item We do not find such an alignment effect with the
      near-IR host morphology. 
      This suggests that the morphology of the extended UV/blue emission is not related
      to the morphology of the stellar component of the host galaxy which dominates in the near-IR.

       \item We observe the same $PA_{host} - \theta_{Pola}$
      behavior for either Radio-Loud or Radio-Quiet objects. This
      observation supports the idea that the UV/blue continuum
      is not entirely due to star formation processes triggered by the radio-jet.

      \item The observed correlation fits a unification
      model where the Type~1/Type~2 dichotomy is essentially determined
      by orientation effects assuming the two component scattering model of Smith et al.
      (\cite{sm04,sm05}). Indeed, depending on the viewing angle to
      the quasar, the polarization would predominantly arise
      from either the equatorial or the polar scattering region giving rise to the observed behavior.

   \end{enumerate}

      In order to strengthen the conclusions and to further investigate
      the correlations presented in this paper, new observations of quasars in the rest-frame
      UV/blue domain are needed, especially of Radio-Quiet objects for
      which few observations at $\lambda_{rest} \le 5000$ \AA~are available.
      The detection of the extended blue/UV continuum region and the measurement
      of its polarization would help to test our interpretation.

\begin{acknowledgements}
      Part of this work was financially supported by a PAI
      (\emph{P\^{o}le} d'Attraction Inter-universitaire) grant (PAI 5/36)
      and a PhD student grant of the Belgian Fund for Scientific Research (F.N.R.S.).
      This work was also
      supported in part by PRODEX Experiment Agreement 90195 (ESA and PPS Science Policy, Belgium).
      Use of ADS and MAST/ESO HST archive. This research has made
      use of the Vizier catalogue access tool and the SIMBAD database, operated at CDS, Strasbourg,
      France.
      B.B. would like to thank V. Chantry for
      the help provided about reduction pipeline for HST NICMOS
      and ACS images, and P. Greenfield and the STScI Help Desk Staff for
      the councils provided for the WFPC2 image reduction.
      We thank the anonymous referee for useful comments and suggestions
      that improved the paper.

\end{acknowledgements}

\longtab{2}{
\begin{longtable}{l c c c r c r r c}
\caption{\label{imres} Results of our modelling of quasars host galaxy.}\\
\hline\hline
Name  & $z$ & QSO Type & Survey & Domain & Host Type & $b/a$ & $PA_{host}$ & Quality\\

\hline
\endfirsthead
\caption{continued.}\\
\hline\hline
Name  & $z$ & QSO Type & Survey & Domain & Host Type & $b/a$ & $PA_{host}$ & Quality\\

\hline
\endhead
\hline
\endfoot

2M~000810+1354   &   0.185   &   RQ &   \emph{Ma03}     &   Vis &   S       &   0.62        &   137     &   1   \\
PG~0026+129     &   0.142   &   RQ &       \emph{Ml01} &   NIR &       E   &       0.86    &       113 &   1   \\
PG~0043+039     &   0.385   &   RQ &   \emph{Bo98}     &   Vis &    U      &   1.00        &   45      &  2  \\
2M~005055+2933   &   0.136   &   RQ &   \emph{Ma03}     &   Vis &   S       &   0.34        &   56      &   1   \\
PG~0052+251     &   0.154   &   RQ &   \emph{Ba97}     &   Vis &   S       &   0.70         &   171     &   1   \\
PHL~909         &   0.171   &   RQ &   \emph{Ba97}     &   Vis &   E       &   0.51        &   129     &   1   \\
2M~010607+2603   &   0.411   &   RQ &   \emph{Ma03}     &   Vis &   S?      &   0.31        &   115     &  1  \\
SDSS~J0123+0044 &   0.399   &   RQ &   \emph{Za06}     &   Vis &   E       &   0.59        &   70      &  1  \\
UM~357          &   0.335   &   RQ &       \emph{Ml01} &   NIR &       E   &       0.75    &       62  &  1  \\
2M~015721+1712   &   0.213   &   RQ &   \emph{Ma03}     &   Vis &   U       &   1.00        &   65      &  2  \\
PKS~0202-76     &   0.389   &   RL &   \emph{Bo98}     &   Vis &   ...        &     ...   &      ... &  3  \\
NAB~0205+02     &   0.155   &   RQ &   \emph{Ba97}     &   Vis &   S       &   0.67        &   140     &  1  \\
2M~022150+1327   &   0.140   &   RQ &   \emph{Ma03}     &   Vis &   S?      &   0.62        &   142     &  1  \\
2M~023430+2438   &   0.310   &   RQ &   \emph{Ma03}     &   Vis &     ... &      ...   &     ...   &  3  \\
0316-346        &   0.265   &   RQ &   \emph{Ba97}     &   Vis &    ...   &   1.00        &   90      &  2  \\
2M~032458+1748   &   0.328   &   RQ &   \emph{Ma03}     &   Vis &   ...    &    ...       &    ...   &  3  \\
2M~034857+1255   &   0.210   &   RQ &   \emph{Ma03}     &   Vis &   U       &   1.00        &   104     &  2  \\
PKS~0440-00     &   0.844   &   RL &       \emph{Ku01} &   NIR &     ...   &     ...   &  ...     &  3  \\
MS~07546+3928   &   0.096   &   RQ &   \emph{Bo98}     &   Vis &   E       &   1.00        &   72      &  2  \\
2M~092049+1903   &   0.156   &   RQ &   \emph{Ma03}     &   Vis &   S       &   0.52        &   24      &  1  \\
SDSS~J0920+4531 &   0.402   &   RQ &   \emph{Za06}     &   Vis &   E       &   0.55        &   160     &  1  \\
PG~0947+396     &   0.206   &   RQ &       \emph{Ml01} &   NIR &       S   &       0.71    &       22  &  1  \\
HE~1029-140     &   0.086   &   RQ &   \emph{Ba97}     &   Vis &   E       &   0.81        &   144     &  1  \\
SDSS~J1039+6430 &   0.402   &   RQ &   \emph{Za06}     &   Vis &   E       &   0.88        &   20      &  1  \\
PG~1048+342     &   0.167   &   RQ &       \emph{Ml01} &   NIR &       S   &       0.65    &       94  &  1  \\
SDSS~J1106+0357 &   0.242   &   RQ &   \emph{Za06}     &   Vis &   S       &   0.66        &   66      &  1  \\
PG~1116+215     &   0.177   &   RQ &   \emph{Ba97}     &   Vis &   E       &   0.81        &   65      &  1  \\
PG~1121+422     &   0.234   &   RQ &       \emph{Ml01} &   NIR &    ...   &   ...    &  ...      &  3  \\
PG~1151+117     &   0.176   &   RQ &       \emph{Ml01} &   NIR &       E   &       0.87    &       149 &  1  \\
PG~1202+281     &   0.165   &   RQ &   \emph{Ba97}     &   Vis &   E       &   0.92        &   117     &  1  \\
PG~1216+069     &   0.334   &   RQ &   \emph{Bo98}     &   Vis &  ...    &    ...      &      ...     & 3  \\
3C~273          &   0.158   &   RL &   \emph{Ba97}     &   Vis &   E       &   0.79        &   61      &  1  \\
2M~125807+2329   &   0.259   &   RQ &   \emph{Ma03}     &   Vis &   E       &   0.72        &   61      &  1  \\
PG~1307+085     &   0.155   &   RQ &   \emph{Ba97}     &   Vis &   U       &   1.00        &   105     &  2  \\
2M~130700+2338   &   0.275   &   RQ &   \emph{Ma03}     &   Vis &   E       &   0.82        &   178     &  1  \\
PG~1309+355     &   0.184   &   RQ &   \emph{Ba97}     &   Vis &   S       &   0.82        &   3       &  1  \\
PG~1322+659     &   0.168   &   RQ &       \emph{Ml01} &   NIR &   ...  &   ...  &   ...  &  3  \\
SDSS~J1323-0159 &   0.350   &   RQ &   \emph{Za06}     &   Vis &   E       &   0.71        &   26      &  1  \\
PG~1352+183     &   0.158   &   RQ &       \emph{Ml01} &   NIR &       E   &       1.00    &       139 &  2  \\
PG~1354+213     &   0.300   &   RQ &       \emph{Ml01} &   NIR &       E   &       0.78    &       166 &  1  \\
PG~1358+043     &   0.427   &   RQ &   \emph{Bo98}     &   Vis &   ...     &      ...    &   ...    &  3  \\
PG~1402+261     &   0.164   &   RQ &   \emph{Ba97}     &   Vis &   S       &   0.61        &   170     &  1  \\
SDSS~J1413-0142 &   0.380   &   RQ &   \emph{Za06}     &   Vis &   E       &   0.68        &   26      &  1  \\
PG~1427+480     &   0.221   &   RQ &       \emph{Ml01} &   NIR &       E   &       1.00    &       158 &  2  \\
PG~1444+407     &   0.267   &   RQ &   \emph{Ba97}     &   Vis &   S       &   0.82        &   48      &  1  \\
2M~145331+1353   &   0.139   &   RQ &   \emph{Ma03}     &   Vis &   S       &   0.57        &   179     &  1  \\
2M~151621+2259   &   0.190   &   RQ &   \emph{Ma03}     &   Vis &   S       &   0.51        &   47      &  1  \\
2M~152151+2251   &   0.287   &   RQ &   \emph{Ma03}     &   Vis &   U       &   1.00        &   2       &  2  \\
2M~154307+1937   &   0.228   &   RQ &   \emph{Ma03}     &   Vis &   U       &   0.82        &   30      &  1  \\
3C~323.1        &   0.266   &   RL &   \emph{Ba97}     &   Vis &   E       &   0.76        &   99      &  1  \\
NAB~1612+26     &   0.395   &   RQ &       \emph{Ml01} &   NIR &       E   &       0.55    &       99  &  1  \\
Mrk~876         &   0.129   &   RQ &       \emph{Ml01} &   NIR &       U   &       1.00    &       131 &  2  \\
2M~163700+2221   &   0.211   &   RQ &   \emph{Ma03}     &   Vis &   S       &   0.29        &   121     &  1  \\
2M~163736+2543   &   0.277   &   RQ &   \emph{Ma03}     &   Vis &   U       &   1.00        &   128     &  2  \\
2M~165939+1834   &   0.170   &   RQ &   \emph{Ma03}     &   Vis &   U       &   0.81        &   61      &  1  \\
2M~170003+2118   &   0.596   &   RQ &   \emph{Ma03}     &   Vis &    ...  &    ...    &   ...    &  3  \\
2M~171442+2602   &   0.163   &   RQ &   \emph{Ma03}     &   Vis &   S       &   0.59        &   169     &  1  \\
2M~171559+2807   &   0.524   &   RQ &   \emph{Ma03}     &   Vis &          &   1.00       &           &   2 \\
KUV~18217+6419  &   0.297   &   RQ &       \emph{Ml01} &   NIR &       S?  &       1.00    &       65  &  2  \\
3C~422          &   0.942   &   RL &       \emph{Ku01} &   NIR &  ...   &    ...     &   ...    &  3  \\
B2-2156+29      &   1.759   &   RL &       \emph{Ku01} &   NIR &      ...  &    ...      &   ...   &  3  \\
2M~222202+1959   &   0.211   &   RQ &   \emph{Ma03}     &   Vis &   E?      &   1.00        &   46      &  2  \\
2M~222554+1958   &   0.147   &   RQ &   \emph{Ma03}     &   Vis &   S       &   0.49        &   141     &  1  \\
PG~2233+134     &   0.325   &   RQ &       \emph{Ml01} &   NIR &       E   &       1.00    &       135 &  2  \\
2M~225902+1246   &   0.199   &   RQ &   \emph{Ma03}     &   Vis &   E       &   0.68        &   69      &  1  \\
2M~230304+1624   &   0.289   &   RQ &   \emph{Ma03}     &   Vis &   U       &   1.00        &   50      &  2  \\
2M~230442+2706   &   0.237   &   RQ &   \emph{Ma03}     &   Vis &   E       &   0.77        &   60      &  1  \\
2M~232745+1624   &   0.364   &   RQ &   \emph{Ma03}     &   Vis &   S       &   0.62        &   76      &  1  \\
2M~234449+1221   &   0.199   &   RQ &   \emph{Ma03}     &   Vis &   ...     &     ...  &   ...   &  3  \\
SDSS~J2358-0009 &   0.402   &   RQ &   \emph{Za06}     &   Vis &   U       &   1.00        &   68      &  2  \\

\end{longtable}
}

\Online
\begin{appendix} 
\section{Details on the observation campaigns used in this study}

In this Appendix, we give some details about each of the observing
campaigns used in this study and briefly described in Sect. \ref{publidata}.

\subsection{Sample of published and reliable data}
\label{reli}
\subsubsection{The visible domain}

\paragraph{\emph{Ci93} (Cimatti et al. 1993)}

     The sample defined in \emph{Ci93} consists in a compilation
     of $z \geq 0.1$ RGs studied in the literature who possess
     polarization measurements in the optical domain. Their total
     sample was made out of 42 RGs from which 8 NLRGs have $z \in [0.5-1.2]$. NLRGs possess an
     obvious host orientation in their rest-frame UV/blue (redshifted to optical)
     continuum images (\emph{Ci93} and references therein).
     The $\Delta \theta_{O}$ given in Tab.~1 of $Ci93$ are the acute offset angle
     between the main orientation of their optical host image (our
     $PA_{host}$) and the optical polarization angle
     $\theta_{Pola}$.

\paragraph{\emph{Di95} (Disney et al. 1995)}

     The observations were made using the Planetary Camera (PC1) of the Wide Field \& Planetary
     Camera2 (WFPC2, Bagget et al. \cite{bag02}) on board of the
     HST. The PC1 is a 800 X 800 pixel camera and has a 0.046 arcsec per pixel resolution.
      Four quasars whose redshift $z$ belongs to [0.25,0.50] were imaged through the broadband F702W filter
     (which corresponds roughly to R band). The host galaxies
     parameters were derived using a two component cross-correlation technique
     which finds the best fit between the observed images and a series of
     galaxy templates (Phillipps \& Boyce \cite{phi92}).

     \paragraph{\emph{Dk96} (de Koff et al. 1996)}

     The sample
     presented here is a part of a more extended sample,
     containing 267 3CR objects. In the sample of \emph{Dk96}, 77 RGs at
     intermediate ($0.1<z<0.5$) redshift were studied with the PC1 camera of the WFPC2 camera through the
     F702W filter. The host galaxy $PA_{host}$ were determined by hand on the images
     by measuring the position angle of the largest extensions of
     the emission regions at a given surface brightness. The
     uncertainty in the derived position angles may thus be as large as $15\degr$ depending on
     the complexity of the objects.

     \paragraph{\emph{Le99} (Lehnnert et al. 1999)}

     43 RLQs selected from the 3CR radio catalog and spanning a redshift range of $z \in [0.3,2.0]$ were
     imaged using the PC1 camera of the WFPC2 instrument on board
     of the HST. The observations were carried out in the F702W filter.
     The $PA_{host}$ were estimated by fitting a series of ellipses on the residuals
     after the substraction of a scaled Point Spread Function (PSF) playing the role of
     the unresolved nucleus. As for some objects the
     ellipticity of the isophotes can vary with radial distance to the center of the quasar, they
     defined the $PA_{host}$ as the position angle of the major axis at a
     surface brightness of 21.5 $m_{F702W}$ arcsec$^{-2}$. They
     estimated the uncertainty on the derived $PA_{host}$ of less than
     $20\degr$ and an uncertainty of less than 0.07 on the
     ellipticity). Note that they do not provide $b/a$ values in
     their Table~3 preventing us of using the $b/a$ criterion for
     these objects. However this is not a problem since no
     $PA_{host}$ are provided for the litigious objects. Moreover,
     we apply a redshift cut ($z<1.20$) in order to avoid the
     presence of tiny hosts for which the $PA_{host}$ determination remains uncertain.

     \paragraph{\emph{Ma99} (Martel et al. 1999)}

     46 3CR radio galaxies were imaged thanks to the PC1 and the Wide Field
     (WF, the resolution of the WFs captors is 0".0996 pixel$^{-1}$) chips of the WFPC2 camera of HST
     through the F702W filter. The target were selected so that
     their redshift belong the range [0,0.1]. They used the Vista command \emph{axes}
     to calculate the flux-weighted position angle of the major axis
     and the ellipticity of the host galaxy.

     \paragraph{\emph{Du03} (Dunlop et al. 2003)}

     For convenience, this sample gathers under the same name the published data by
     Dunlop et al. \cite{du03} and McLure et al. \cite{ml99}
     because they are both taken from the same HST survey.
     33 objects (RQQs, RLQs and RGs) with $z \in [0.1,0.25]$ were observed with the
     WF2 camera of the WFPC2 instrument through the F675W
     filter (corresponding roughly to R band).
     The parameters of the host
     galaxies were determined by fitting an analytical
     model to the PSF-subtracted images (McLure et al. \cite{ml00}).

     It is wise to pay attention to the fact that the values given in
     Table~3 of \emph{Du03} do not fit the $PA_{host}$ definition adopted
     here, contrary to what is claimed in the caption, but are determined with respect to the $y-axis$ of the images.
     We have thus corrected these values applying a correction (the "ORIENTAT" parameter
     contained in the header of the observation files) taking account of
     the orientation of the HST with respect to the plane of the sky.

     \paragraph{\emph{Fl04} (Floyd et al. 2004)}

      17 quasars with $z \in [0.29,0.34]$ (both Radio-Loud and Radio-Quiet) were observed
      through the WF camera of the WFPC2 instrument on board of the
      HST. The observations were carried in two photometric bands
      (the broad F814W and narrow F791W filters roughly corresponding
      to I band) in order to avoid the contamination by strong emission lines
      depending on the redshift of the object. The procedure used to
      derive the host galaxies parameters is the same as used in
      \emph{Du03}.

      It is clearly stated that the position angles of the
      studied hosts are given anticlockwise from the vertical in the images. We
      thus corrected these values by the same way as those of
      \emph{Du03} and derived the corresponding $PA_{host}$.

      \paragraph{\emph{Mc04} (McLure et al. 2004)} 41 radio
      galaxies were observed in the I-band (F785LP filter) using
      the WF3 chip of the WFPC2. The redshift of the sample
      is comprised between $0.4<z<0.6$. The technique used to
      derive the host galaxies parameters is the same
      two-dimensional modelling process as the one used by
      \emph{Du03}.

   \subsubsection{The near-IR domain}

      \paragraph{\emph{Ta96} (Taylor et al. 1996)}

      44 quasars were observed in the photometric K-band
      ($\lambda_{cent} \sim 2.2 \mu m$) using the IRCAM camera of
      the 3.9 m United Kingdom Infrared Telescope (UKIRT). The
      sample consists of RLQs, RQQs and RGs having a redshift
      around 0.2. Let us notice that the majority of these objetcs
      have been imaged in visible wavelengths by \emph{Du03}. The
      parameters of the host galaxies were obtained using a fully
      two dimensional modelling procedure.

      \paragraph{\emph{Pe01} (Percival et al. 2001)}

       13 objects spanning a redshift range of [0.26,0.46] were observed in
       the K-band using the IRCAM3 camera of the UKIRT telescope.
       As the observations were carried out from the ground for
       relatively distant objects, the host galaxies are
       completely hidden by the PSF wings of the nuclear
       component. Each image was then deconvolved to subtract a nuclear PSF and recover the
       weak glare of the underlying host galaxy. The
       PSF-subtracted images were then modelled by a two
       dimensional brightness analytical galaxy profile. The
       orientation of the host galaxy (called $\alpha$ by \emph{Pe01}) is given in their Table 4. It is expressed in radians and related to the $PA_{host}$ definition adopted in our
       paper following $PA_{host} = \alpha$ (in degrees) $+ 90$ (or $PA_{host} = \alpha$ $- 90$
       if $\alpha$ (in degrees) is smaller than $90\degr$).

       \paragraph{\emph{Gu06} (Guyon et al. 2006)}

       32 quasars at $z<0.3$ selected from the Palomar-Green Bright Quasar
       Survey were imaged in the near infrared domain (H-band) using
       adaptive optics (AO) imaging with the Gemini 8.2 m and
       Subaru 8.1 m telescopes on Mauna Kea. The images were PSF-subtracted
       and residuals were modelled using analytical galaxy models.

       \paragraph{\emph{Ve06} (Veilleux et al. 2006)}

       33 quasars were observed in the H-band using the Near
       Infrared Camera and Multi Object Spectrometer (NICMOS)
       instrument on board of the HST. The sample consists of $z<0.3$
       luminous late-stage galactic mergers. The removal of the PSF and the structural
       parameters of the underlying hosts were determined using
       the GALFIT (Peng et al. \cite{peng02}) algorithm.

\subsection{ Sample of \emph{new $PA_{host}$ data}}
\label{newly}
\subsubsection{The visible domain}

 \paragraph{\emph{Ba97} (Bahcall et al. 1997)}

          The sample observed by \emph{Ba97} contains 20 quasars $z$ $<$
          0.3 (both RQQs and RLQs) imaged with the WF3 camera of the WFPC2
          instrument through the F606W filter (whose spectral characteristics
          are $\lambda_{cent} \sim 5900$~\AA~and $\Delta \lambda \sim 1500$~\AA).
          Due to the redshift of the targets, powerful
          emission lines such as $[OIII]$ and $H_{\alpha}$ are
          included in the image. The objects in the sample
          come from the 1991 V\'eron-Cetty \& V\'eron catalog and were
          selected such as $z \leq 0.2$, $M_{V} < -22.9$ et $|b| > 35\degr$
          (where $b$ stands for the galactic latitude). Several
          objects present in the sample were also part of the sample of
          \emph{Du03} and were useful to assess our $PA_{host}$
          determination method (cf. Sect.~\ref{compa}).

  \paragraph{\emph{Bo98} (Boyce et al. 1998)}

      7 quasars were imaged with the WFPC2 instrument on board of
      the HST. Six of them were observed through the PC1 camera
      and the last one (PG0043+039) was observed using the WF chip. All the
      observations were taken through the F702W filter. The
      objects span a redshift range of $0.09<z<0.43$ and contain
      both RLQs and RQQs.

    \paragraph{\emph{Ma03} (Marble et al. 2003)}

       29 RQQs spanning $0.136 < z < 0.596$ were observed with the
       PC1 camera of the WFPC2 instrument through the F814W. The targets were selected from the Two Micron All
       Sky Survey (2MASS) essentially according to color and galactic latitude criterion.
       Note that the 2MASS sample contains reddened Type~1
       quasars which, in the quasar unification scheme, can be
       interpreted as normal quasars, but seen close to the
       plane of an obscuring torus.
       The 2MASS objects have been classified as QSOs based on the
       fact that they are as luminous at $2.2 \mu m$ as QSOs in
       other samples (e.g. PG objects of comparable $z$).

      \paragraph{\emph{Za06} (Zakamska et al. 2006)}

         9 Type~2 Radio-Quiet quasars at $z \in[0.2,0.4]$ were imaged through the Wide
         Field Channel (WFC) of the ACS instrument (Pavlovsky et al. \cite{pa05})
         in three photometric band ($Yellow$, $Blue$ and $UV$ cf.
         \emph{Za06}). The resolution of the WFC is similar to the resolution of the PC1 camera of the WFPC2 instrument.
         Their targets were selected from the
         sample of Type~2 active galactic nuclei candidates unveiled by Zakamska et al. (\cite{za03}) and Hao et al.
         (\cite{ha05}).

    \subsubsection{The near-IR domain}

    \paragraph{\emph{Ml01} (McLeod et al. 2001)}

    16 RLQs were imaged in the near-infrared domain (with the F160W
    filter corresponding approximately to H band), using the
    NIC2 camera ($0.075" pixel^{-1}$ resolution) of NICMOS on board of
    the HST. The objects have $z<0.4$ and were imaged in the
    MULTIACCUM mode (Schultz et al. \cite{shu05}) which records
    data in a series of increasingly long nondestructive readouts,
    providing high dynamic unsaturated images.

    \paragraph{\emph{Ku01} (Kukula et al. 2001)}

     20 high redshift ($z \in [0.83,2.01]$) quasars were imaged using the NICMOS NIC1
     camera (resolution of $\sim 0.043" pixel^{-1}$). Due to the
     large redshift range, two filters were used to image the
     quasars near the same rest-frame wavelength
     and to avoid the presence of strong emission lines such as $[OIII]$ and $H_{\beta}$.
     Thus the F110M filter (roughly J-band) was used to image the quasars in the $z \in [0.83,1.00]$ range,
     and the F165M (roughly H-band)
     filter to image the quasars in the $z \in [1.67,2.01]$ range.

\end{appendix}

\begin{appendix} 
\section{Comparison with previous studies}
\label{compapp}

In this Appendix we give some details on the comparison of the
results obtained with the modelling procedure adopted in this
paper and the results derived in published papers (listed in
Sect.~\ref{publidata}) as five objects of our \emph{new
$PA_{host}$ data} were previously studied. Hereafter, for each
object of our sample being also part of another study where host
parameters were determined, we give a brief description of the
object and compare the derived $b/a$ and $PA_{host}$ obtained. In
general we note a good agreement between both sets of parameters.

         \paragraph{PG 0052+251}
         The galaxy of this RQQ from the sample of \emph{Ba97}
         consists in a spiral host with two prominent arms
         presenting several knots identified as $HII$ regions.
         This object was already studied in two papers : Bahcall et al.
         \cite{ba96} (\emph{Ba96}) and
         \emph{Du03} (who noted that the galaxy profile was best modelled by a spheroidal component
         rather than a disk-like one). The results of these
         studies ($PA=173\degr$ et $b/a=0.66$ for \emph{Ba96} and $PA=175\degr$ et $b/a=0.61$ for \emph{Du03}) are
         very close to the one we obtain ($PA=171\degr$ and $b/a=0.70$).

         \paragraph{PHL 909}
         This RQQ host observed by \emph{Ba97} was
         characterized in two previous studies : \emph{Ba96} and McLure et al. \cite{ml99} (referred as \emph{Du03} in our study). The host parameters of the elliptical galaxy we obtained
         with our method ($b/a$ = 0.51 and $PA = 129\degr$) are similar to those found by both
         studies (\emph{Ba96} found $b/a$ = 0.5 and $PA = 138\degr$ and \emph{Du03} $b/a$ = 0.61 and $PA = 131\degr$).

          \paragraph{UM 357}
          This RQQ from the sample of NIR studied QSOs \emph{Ml01}
          has been previously studied from the ground in a similar waveband by
          \emph{Pe01}. The derived parameters using our technique
          ($PA=62\degr$ et $b/a=0.75$) are close to those obtained by
          \emph{Pe01} ($PA=56\degr$ and $b/a=0.66$).

         \paragraph{PG 1354+213}
         This RQQ was studied by
         \emph{Pe01} in the K-band. In our sample,
         the observations of this object comes from the \emph{Ml01}
         H-band survey. As the two images were taken in Near-IR, we
         can generally expect a good match between the K-band and H-band galaxy parameters.
         This is effectively the case for this object for which \emph{Pe01} gives
         $PA=169\degr$ and $b/a=0.89$ while we obtain $PA=166\degr$ and $b/a=0.78$.

         \paragraph{PG 2233+134}
         The host galaxy parameters were formerly derived by
         \emph{Pe01} in the near-IR. They obtained $PA=178\degr$ and $b/a=0.8$. These
         parameters are quite different from the one we obtained ($PA=135\degr$ and $b/a=0.89$)
         using the \emph{Ml01} images.
         However, as mentioned in the Table~\ref{imres} our
         results do not satisfy the quality criterion defined
         in the Sect.~\ref{compa}.

\end{appendix}


\begin{thebibliography}{}

  \bibitem[1983]{anto83} Antonucci, R. 1983,
  \nat, 303, 158

  \bibitem[1985]{anto85} Antonucci, R. \& Miller, J. 1985,
  \apj, 93, 785

  \bibitem[1993]{anto93} Antonucci, R. 1993,
  \araa, 1993.31, 473

  \bibitem[2002]{bag02} Bagget, S. et al.
  2002, WFPC2 Data Handbook, version 4.0 (Baltimore : STScI)

  \bibitem[1996]{ba96} Bahcall, J.N., Kirhakos, S. \& Schneider, D.P. 1996,
  \apj, 457, 557

  \bibitem[1997]{ba97} Bahcall, J.N., Kirhakos, S. \& Saxe, D.H. 1997,
  \apj, 479, 642

  \bibitem[1989]{bar89} Barthel, P.D. 1989,
  \aj, 336, 606

  \bibitem[1990]{beri90} Berriman, G., Schmidt, G.D., West, S.C. \& Stockman, H.S. 1990,
  \apjs, 74, 869

  \bibitem[1998]{be98} Best, P.N., Longair, M.S. \& Röttgering, J.A. 1998, \mnras, 295, 549

  \bibitem[2002]{bir02} Biretta, J.A., Martel, A.R., McMaster, M., et al. 2002,
  \nar, 46, 181
  
  \bibitem[1994]{bo94} Bogers, W.J., Hes, R., Barthel, P.D. \& Zensus, J.A. 1994,
  \aap, 105,91

  \bibitem[1998]{bo98} Boyce, P.J., Disney, M.J., Blades, J.C., et al. 1998,
  \mnras, 298, 121
 
  
  \bibitem[1994]{bri94} Bridle, A.H., Hough, D.H., Lonsdale, C.J., Burns, J.O. \& Laing, R.A. 1994,
  \aj, 108, 766

  \bibitem[1987]{cha87} Chambers, K.C., Miley, G.K. \& van Breugel, W. 1987,
  \nat, 329, 604

 
  \bibitem[1993]{cima93} Cimatti, A., di Serego-Alighieri, S., Fosbury, R.A.E., Salvati, M. \& Taylor, D. 1993, \mnras, 264, 421

  \bibitem[1999]{coh99} Cohen, M.H., Ogle, P.M., Tran, H.D.,
  Goodrich, R.W. \& Miller, J.S. 1999,
  \aj, 118, 1963

  \bibitem[1996]{dk96} de Koff, S., Baum, S.A., Sparks, W.B., et al. 1996,
  \apjs, 107, 621 
  
  
  \bibitem[1996]{dey96} Dey, A., Cimatti, A., van Breugel, W., Antonucci, R. \& Spinrad, H. 1996,
  \apj, 465, 157

  \bibitem[1989]{di89} di Serego Alighieri, S., Fosbury, R.A.E. 1993, Quinn P.J.
  \& Tadhunter, C.N. 1989, \nat, 341, 307

  \bibitem[1993]{di93} di Serego Alighieri, S., Cimatti, A. \&
  Fosbury, R.A.E. 1993, \apj, 404, 584

  \bibitem[1995]{di95} Disney, M.J., Boyce, P.J., Blades, J.C., et al. 1995,
  \nat, 376, 150
  
  \bibitem[1993]{du93}Dunlop, J.S. \& Peacock, J.A. 1993, \mnras, 263, 936

  \bibitem[2003]{du03} Dunlop, J.S., McLure, R.J., Kukula, M.J., et al. 2003,
  \mnras, 340, 1095
  
  
  \bibitem[2004]{fl04} Floyd, D.J.E., Kukula, M.J., Dunlop, J.S., et al. 2004,
  \mnras, 355, 196
 
  \bibitem[2007]{goo07} Goosmann, R.W. \& Gaskell, C.M. 2007,
  \aap, 465, 129

  \bibitem[1984a]{go84a} Gower, A.C. \& Hutchings, J.B. 1984a,
  \pasp, 96, 19

  \bibitem[1984b]{go84b} Gower, A.C. \& Hutchings, J.B. 1984b,
  \aj, 89,1658

  \bibitem[2006]{gr06} Grandi, P., Malaguti, G. \& Fiocchi, M. 2006,
  \apj, 642, 113

  \bibitem[2006]{gu06} Guyon, O., Sanders, B.D., Stockton, A. 2006,
  astro-ph/0605079v1

  \bibitem[2005]{haa05} Haas, M., Siebenmorgen, R., Shultz, B., Kr\"{u}gel, E. \& Chini, R. 2005,
  \aap, 442, L39

  \bibitem[2005]{ha05} Hao, L., Strauss, M.A., Tremonti, C.A., et al. 2005,
  \aj, 129, 1783
  
  \bibitem[1983]{hi83} Hintzen, P., Ulvestad, J. \& Owen, F. 1983,
  \aj, 88, 709

  \bibitem[1999]{hu99} Hurt, T., Antonucci, R., Cohen, R., Kinney, A. \& Krolik, J. 1999,
  \apj, 514, 579

  \bibitem[2005]{hu05} Hutsem\'ekers, D., Cabanac, R., Lamy, H. \& Sluse, D. 2005,
  \aap, 441, 915

  \bibitem[1997]{ja97} Jackson, N. \& Rawlings, S. 1997,
  \mnras, 286, 241

  \bibitem[2004]{ja04} Janhke, K., Sanchez, S.F., Wisotzki, L.,
  et al. 2004, \aj, 614, 568
 
  \bibitem[1989]{ke89} Kellerman, K.I., Sramek, R., Schmidt, M., Shaffer, D.B. \& Green, R. 1989,
  \aj, 98, 1195

  \bibitem[2007]{ki07} King, A.R. \& Pringle, J.E. 2007, \mnras, 377, L25

  \bibitem[2000]{ki00} Kinney, A.L., Schmitt, H.R., Clarke, C.J., et al. 2000, \apj, 537, 152
  
  \bibitem[2002]{ko02} Koekemoer, A.M., Fruchter, A.S., Hook,
  R.N. \& Hack, W. 2002,
  HST Calibration Workshop, eds. S. Arribas, A.M. Koekemoer \& B. Whitmore (Baltimore : STScI)

  \bibitem[2004]{kot04} Kotilainen, J.K. \& Falomo, R. 2004,
  \aap, 424, 107

  \bibitem[2004]{kr04} Krist, J., Hook., R. 2004,
  http://www.stsci.edu/software/tinytim

  \bibitem[2001]{ku01} Kukula, M.J., Dunlop, J.S., McLue, R.J., et al. 2001,
  \mnras, 326, 1533
  
  \bibitem[1987]{law87} Lawrence, A. 1987,
  \pasp, 99, 309

  \bibitem[1999]{le99} Lehnert, M.D., Miley, G.K., Sparks, W.B., et al. 1999,
  \apjs, 123, 351
 
  \bibitem[1994]{li94} Lister, M.L., Gower, A.C. \& Hutchings, J.B. 1994,
  \aj, 108, 821

  \bibitem[2000]{li00} Lister, M.L. \& Smith, P.S. 2000,
  \aj, 541, 66

  \bibitem[1987]{ma87} McCarthy, P.J., van Breugel, W.J.M.,
  Spinrad, H. \& Djrgovski, S. 1987,
  \apj, 321, L29

  \bibitem[1997]{ma97} McCarthy, P.J., Miley, G.K., de Koff, S.,
  et al. 1997, \apjs, 112, 415
  
  \bibitem[2001]{mc01} McLeod, K.K. \& McLeod, B.A., 2001,
  \apj, 546, 794

  \bibitem[2004]{mc04} McLure, R.J., Willott, C.J., Jarvis, M.J.,
  et al. 2004, \mnras, 351, 347
 
  \bibitem[1999]{ml99} McLure, R.J., Kukula, M.J., Dunlop, J.S., et al. 1999,
  \mnras, 308, 377
  
  \bibitem[2000]{ml00} McLure, R.J., Dunlop, J.S. \& Kukula, M.J. 2000,
  \mnras, 318, 693

  \bibitem[1998]{ma98} Magain, P., Courbin, F. \& Sohy, S. 1998,
  \apj, 449, 472

  \bibitem[2003]{ma03} Marble, A.R., Hines, D.C., Schmidt, G.D., et al. 2003,
  \apj, 590, 707
  
  \bibitem[1999]{ma99} Martel, A.R., Baum, S.A., Sparks, W.B., et al. 1999,
  \apjs, 121, 81
 
  \bibitem[1984]{ow84} Owen, F.N. \& Puschell, J.J., 1984,
  \aj, 89, 932

  \bibitem[2005]{pa05} Pavlovsky, C. et al. 2005,
  ACS Data Handbook, version 4.0 (Baltimore : STScI)

  \bibitem[2002]{peng02} Peng, C., Ho, L.C., Impey, C.D. \& Rix, H.-W. 2002,
  \aj, 124, 266

  \bibitem[2001]{pe01} Percival, W.J., Miller, L., McLure, R.J. \& Dunlop,
  J.S. 2001,
  \mnras, 322, 843

  \bibitem[1992]{phi92} Phillipps, S., Boyce, P.J. 1992,
  \mnras, 256, 673

  \bibitem[2002]{pi02} Pian, E., Falomo, R., Hartman, R.C.,
  et al. 2002, \aap, 392, 407
  
  \bibitem[1993]{pri93} Price, R., Gower, A.C., Hutchings, J.B.,
  et al. 1993, \apjs, 86, 365
 
  
  \bibitem[1979]{re79} Readhead, A.C.S., Pearson, T.J., Cohen,
  M.H., Ewing, M.S. \& Moffet, A.T. 1979, \apj, 231, 299

  \bibitem[1995]{re95} Reid, A., Shone, D.L., Akujor, C.E.,
  et al. 1995, \aap, 110, 213
 
  \bibitem[1992]{ri92} Rigler, M.A., Lilly, S.J. Stockton, A.,
  Hammer, F. \& Le F\`{e}vre, O. 1992,
  \apj, 385, 61

  \bibitem[1983]{rudy83} Rudy, R.J., Schmidt, G.D., Stockman, H.S.
  \& Moore, R.L. 1983,
  \aj, 271, 59

  \bibitem[1985]{rusk85} Rusk, R. \& Seaquist, E.R., 1985,
  \aj, 90, 30

  \bibitem[2004]{san04} Sanchez, S.F., Jahnke, K., McIntosh, D.H.,
  et al. 2004, \apj, 614, 586
  
  \bibitem[1983]{shm83} Schmidt, M. \& Green, R.F. 1983,
  \apj, 269, 352

   \bibitem[2001]{sch01} Schmitt, H.R., Ulvestad, J.S., Kinney, A.L., et al. 2001, ASPC, 249, 230
  
  \bibitem[2005]{shu05} Schultz, A. et al. 2005,
  NICMOS Instrument Handbook, version 8.0 (Baltimore : STScI)

  \bibitem[1968]{se68} S\'ersic, J.L. 1968,
  Atlas de Galaxias Australes, Observatorio Astronomico, C\'ordoba

  \bibitem[2004]{si04} Siebenmorgen, R., Freudling, W., Kr\"{u}gel, E., Haas, M. 2004,
  \aap, 421, 129

 \bibitem[2002]{smi02} Smith, J.E., Young, S., Robinson, A., et al. 2002, \mnras, 335, 773
 
  \bibitem[2004]{sm04} Smith, J.E., Robinson, A., Alexander, et al. 2004,
  \mnras, 350, 140
  
  \bibitem[2005]{sm05} Smith, J.E., Robinson, A., Young, S., Axon, D.J. \& Corbett, E.A. 2005,
  \mnras, 359, 846

  \bibitem[2002]{sm02} Smith, P.S., Schmidt, G.D., Hines, D.C., Cutri, R.M. \& Nelson, B.O. 2002,
  \apj, 569, 23

  \bibitem[1991]{spe91} Spencer, R.E., Schilizzi, R.T., Fanti, C.,
  et al. 1991, \mnras, 250, 225
  
  \bibitem[1979]{sto79} Stockman, H.S., Angel, R.J.P. \& Miley, G.K. 1979,
  \apj, 227, L55

  \bibitem[1998]{tad98} Tadhunter, C.N., Morganti, R., Robinson,
  A., et al. 1998, \mnras, 298, 1035
  
  \bibitem[2002]{tad02} Tadhunter, C.N., Dickson, R., Morganti,
  R., et al. 2002, \mnras, 330, 977
  
  \bibitem[1996]{ta96} Taylor, G.L., Dunlop, J.S., Hughes, D.H. \& Robson, I.E. 1996,
  \mnras, 283, 930

  \bibitem[1988]{toma88} Thompson, I.A. \& Martin, P.G. 1988,
  \apj, 330, 121

  \bibitem[1995]{urry95} Urry, M.C. \& Padovani, P. 1995,
  \pasp, 107, 803

  \bibitem[2001]{vb01} van Bemmel, I. M. \& Bertoldi, F. 2001,
  \aap, 368, 414

  \bibitem[2006]{ve06} Veilleux, S., Kim, D.C., Peng, C.Y., et al. 2006,
  \apj, 643, 707
  
  \bibitem[2006]{ver06} V\'eron-Cetty, M.P. \& V\'eron, P. 2006,
  \aap, 455, 773

  \bibitem[2003]{za03} Zakamska, N.L., Strauss, M.A., Krolik, J.H., et al. 2003,
  \aj, 126, 2125
  
  \bibitem[2005]{za05} Zakamska, N.L., Schmidt, G.D., Smith, P.S.,
  et al. 2005, \aj, 129, 1212
  
  \bibitem[2006]{za06} Zakamska, N.L., Strauss, M.A., Krolik, J.H., et al. 2006,
  \aj, 132, 1496
  


\end{thebibliography}
\end{document}